%% file: main.tex
\documentclass[10pt,conference]{IEEEtran}
\IEEEoverridecommandlockouts

\usepackage{tikz}
\usetikzlibrary{arrows.meta,
                chains,
                positioning,
                shapes.geometric,
                shapes.symbols,
                decorations.pathreplacing,
                calligraphy
                }

\usepackage{amsmath,amsfonts,amsthm}
\usepackage{algpseudocode,algorithm}
\usepackage{tabularx,colortbl}
\usepackage{adjustbox}
\usepackage{comment}
\usepackage{pifont}
\usepackage{fancybox,graphicx}
\usepackage{textcomp}
\usepackage{booktabs}
\usepackage{diagbox}
\usepackage{subcaption}
\usepackage{multirow, multicol}
\usepackage{soul}
\usepackage{hyperref}
\usepackage{xspace}
\usepackage{makecell}
\usepackage{enumitem}
\usepackage[absolute,overlay,showboxes]{textpos}

\usepackage{MnSymbol}

\newcommand{\RN}[1]{%
  \textup{\uppercase\expandafter{\romannumeral#1}}%
}
\usepackage{fontawesome}
\usepackage{xcolor}
\usepackage{array,colortbl,xcolor}

\usepackage{scalerel}

\def\BibTeX{{\rm B\kern-.05em{\sc i\kern-.025em b}\kern-.08em
    T\kern-.1667em\lower.7ex\hbox{E}\kern-.125emX}}
    
\usepackage{tcolorbox}
\usepackage{proof}
\definecolor{darkspringgreen}{rgb}{0.09, 0.45, 0.27}

\definecolor{green}{rgb}{0.0, 0.65, 0.31}
\definecolor{jasper}{rgb}{0.84, 0.23, 0.24}
\definecolor{redniki}{rgb}{0.99, 0.40, 0.40}
\definecolor{cookie}{rgb}{1.0, 0.86, 0.35}
\definecolor{almond}{rgb}{0.94, 0.87, 0.8}
\definecolor{cottoncandy}{rgb}{1.0, 0.74, 0.85}
\definecolor{lightapricot}{rgb}{0.99, 0.84, 0.69}
\definecolor{eclipse}{RGB}{127,0,85}

\newcommand{\boldparagraph}[1]{\vskip 0.05in\noindent\textbf{#1.}}

\newcommand{\italicparagraph}[1]{\vskip 0.05in\noindent\textit{(#1).}}

\newcommand{\imply}[2]{#1 \; \sleeckeyword{imply} \; #2}

\newcommand{\hyp}[2]{#1 \; \sleeckeyword{hypernym} \; #2}

\newcommand{\ops}[2]{#1 \; \sleeckeyword{opposite} \; #2}

\newcommand{\eq}[2]{#1 \; \sleeckeyword{equal} \; #2}

\newcommand{\hb}[2]{#1 \; \sleeckeyword{happenBefore} \; #2}

\newcommand{\negate}[1]{\sleeckeyword{not}~(#1)}

\newcommand{\me}[2]{#1 \; \sleeckeyword{mutualExc} \; #2}

\newcommand{\con}[2]{#1 \; \sleeckeyword{isContradictoryWith} \; #2}

\newcommand{\fob}[2]{#1 \; \sleeckeyword{forbids} \; #2}

\newcommand{\ind}[2]{#1 \; \sleeckeyword{indu} \; #2}

\newcommand{\wuntil}[3]{\sleeckeyword{when} \; #1  \; \sleeckeyword{then} \; #2 \; \sleeckeyword{until} \; #3}

\newcommand{\wfor}[3]
{\sleeckeyword{when} \; #1  \; \sleeckeyword{then} \; #2 \; \sleeckeyword{for} \; #3}

\theoremstyle{definition}
\newtheorem{definition}{Definition}
\newtheorem{remark}{Remark}
\newtheorem{example}{Example}

\definecolor{turquoisegreen}{rgb}{0.63, 0.84, 0.71}
\definecolor{tuftsblue}{rgb}{0.28, 0.57, 0.81}
\definecolor{atomictangerine}{rgb}{1.0, 0.6, 0.4}
\definecolor{lightapricot}{rgb}{0.99, 0.84, 0.69}
\definecolor{babypink}{rgb}{0.96, 0.76, 0.76}
\definecolor{mistyrose}{rgb}{1.0, 0.89, 0.88}
\definecolor{moccasin}{rgb}{0.98, 0.92, 0.84}
\definecolor{teagreen}{rgb}{0.82, 0.94, 0.75}
\definecolor{lavender}{rgb}{0.9, 0.9, 0.98}
\definecolor{mygray}{rgb}{0.5,0.5,0.5}
\definecolor{mymauve}{rgb}{0.1,0.2,0.7}
\definecolor{olivegreen}{cmyk}{.6,.4,0.8,0}
\definecolor{eclipse}{RGB}{127,0,85}
\definecolor{darkpastelpurple}{rgb}{0.59, 0.44, 0.84}

\newcommand{\eclipsecolor}[1]{\textcolor{eclipse}{#1}}

\sethlcolor{lightapricot}

\makeatletter
\algrenewcommand\ALG@beginalgorithmic{\scriptsize}
\makeatother

\newcommand{\sleec}[1]{{\scriptsize\fontfamily{qcr}\selectfont#1}}
\newcommand{\sleecT}[1]{{\footnotesize\fontfamily{qcr}\selectfont#1}}
\newcommand{\sleeckeyword}[1]{\eclipsecolor{\textbf{#1}}}

\newcommand{\WFI}{WFI\xspace}
\newcommand{\SAN}{LLM-San\xspace}

\newcommand{\fol}{FOL$^*$\xspace}
\newcommand{\dsl}{\texttt{SLEEC}\xspace}

\newcommand{\tool}{{\texttt{LEGOS-SLEEC}}\xspace}
\newcommand{\approach}{{\texttt{RAINCOAT}}\xspace}
\newcommand{\quoted}[1]{``{#1}''}

\newcommand{\andS}{\textbf{and}}
\newcommand{\orS}{\textbf{or}}
\newcommand{\notS}{\textbf{not}\xspace}

\newcommand{\within}[2]{#1\, \textbf{within} \,#2}
\newcommand{\rulesyntax}[2]{\textbf{when}\, #1 \, \textbf{then}  \,#2}
\newcommand{\factsyntax}[2]{\textbf{exists}\, #1 \, \textbf{while}  \,#2}
\newcommand{\lang}[1]{\mathcal{L}(#1)}

\newcommand{\events}{E}
\newcommand{\measures}{M}
\newcommand{\ruleset}{\textit{Rules}}
\newcommand{\fos}{\sigma}

\newcommand{\term}{t}
\newcommand{\constant}{c}
\newcommand{\measure}{m}
\newcommand{\event}{e}
\newcommand{\obg}{ob}
\newcommand{\cobg}{cob}
\newcommand{\otherwise}{\textbf{ otherwise }}
\newcommand{\sand}{\textbf{ and }}

\newcommand{\prop}{p}

\newcommand{\dobg}{\bigvee_{\cobg}}
\newcommand{\srule}{r}
\newcommand{\fact}{f}
\newcommand{\facts}{Facts}
\newcommand{\eventocc}[1]{\mathcal{E}_{#1}}
\newcommand{\measureassign}[1]{\mathbb{M}_{#1}}
\newcommand{\timestamp}[1]{\delta_{#1}}

\begin{document}

\setlength{\abovedisplayskip}{1pt}
\setlength{\belowdisplayskip}{1pt}

\newcommand{\affilUT}{\affiliation{%
  \institution{University of Toronto}
  \city{Toronto}
  \country{Canada}
}}
\newcommand{\affilUY}{\affiliation{%
  \institution{University of York}
  \city{York}
  \country{UK}
}}
\newcommand{\affilSC}{\affiliation{%
  \institution{Smith College}
  \city{Northampton}
  \country{USA}
}}
\newcommand{\affilUB}{\affiliation{%
  \institution{University of Brasilia}
  \city{Brasilia}
  \country{Brazil}
}}




\title{Normative Requirements Operationalization with Large Language Models}
\author{
    \IEEEauthorblockN{Nick Feng\IEEEauthorrefmark{1}, Lina Marsso\IEEEauthorrefmark{1}, Sinem Getir Yaman\IEEEauthorrefmark{2}, Isobel Standen\IEEEauthorrefmark{2}}
    \IEEEauthorblockN{Yesugen Baatartogtokh\IEEEauthorrefmark{1}, Reem Ayad\IEEEauthorrefmark{1}, Victória Oldemburgo de Mello\IEEEauthorrefmark{1}, Beverley Townsend\IEEEauthorrefmark{2}}
    \IEEEauthorblockN{Hanne Bartels\IEEEauthorrefmark{1}, Ana Cavalcanti\IEEEauthorrefmark{2}, Radu Calinescu\IEEEauthorrefmark{2}, Marsha Chechik\IEEEauthorrefmark{1}}
    \IEEEauthorblockA{\IEEEauthorrefmark{1} University of Toronto, Toronto, Canada
    \\\{fengnick,lmarsso,chechik\}@cs.toronto.edu}
    \IEEEauthorblockA{\IEEEauthorrefmark{2} University of York, York, UK
    \\\{sinem.getir.yaman,cavalcanti,calinescu\}@york.ac.uk}
}
\vspace{-0.6in}

\maketitle 

\thispagestyle{plain} 
\pagestyle{plain} 

\begin{abstract}  
Normative non-functional requirements specify constraints that a system must observe in order 
to avoid violations of social, legal, ethical, empathetic, and cultural norms. As these requirements are typically defined by non-technical system stakeholders with different expertise and priorities (ethicists, lawyers, social scientists, etc.), ensuring their well-formedness and consistency is very challenging. 
Recent research has tackled this challenge using a domain-specific language to specify normative requirements as rules whose consistency can then be analysed with formal methods.
In this paper, we propose 
a complementary approach that uses Large Language Models 
to extract semantic relationships between abstract representations of system capabilities.  These relations, which are often assumed implicitly by non-technical stakeholders (e.g., based on common sense or domain knowledge), are then used to enrich  
the automated reasoning techniques 
for eliciting and analyzing the consistency of normative requirements. 
We show the effectiveness of 
our approach to normative requirements elicitation and operationalization 
through a range of real-world case studies.

\end{abstract}

\input{Sections/introduction}

\input{Sections/background}
\input{Sections/sanitization}
\input{Sections/approach}
\input{Sections/evaluation}
\input{Sections/related-work}
\input{Sections/conclusion}

\bibliographystyle{splncs04}
\bibliography{refs}
\clearpage
\newpage
\appendix
\input{Sections/Appendix/appendix}

\end{document}

%% file: Sections/introduction.tex

\section{Introduction}

\textbf{Context: normative requirements (NRs).} Software systems are increasingly capable of performing daily tasks alongside humans, as important components of our infrastructure. Therefore, ensuring the \emph{responsible} integration of these systems into our society is becoming crucial. This includes preventing them from violating social, legal, ethical, empathetic, and cultural~(SLEEC) norms~\cite{townsend-et-al-2022,Getir-Yaman-et-al-23,Getir-Yaman-et-al-23b}. As such, it is important to elicit \emph{normative} \emph{requirements}~(NRs), which specify the permissible range of a system's behaviours. 
For instance, a legal NR requirement for an assistive robot, would be \textit{``Ra: Users of care services should be treated in a way that ensures their privacy and autonomy''}~\cite{care2012state}.
NR requirements encompass both functional and non-functional aspects and are delineated with respect to time. These requirements can either be specified at a high level, to ensure that the system has adequate capabilities, or  at a lower-level, enabling direct operationalization.
For instance, Ra is a high-level requirement.  On the other hand, the lower-level autonomy and privacy requirement \textit{``Rb: When the user tells the robot to open the curtains then the robot should open the
curtains, unless the user is ‘undressed’ in which case the robot does not open the curtains%
''} is a direct operationalization of Ra~\cite{townsend-et-al-2022}. 
Rb is described with respect to the robot's capabilities to accept requests, identify whether the user is undressed, and open the curtains. NRs can include requirements of both types. 
Due to the nature of the NRs, their elicitation involves non-technical stakeholders (e.g., ethicists, lawyers), each with different and potentially conflicting jargon, goals, priorities, and responsibilities. So, it is particularly important to support the non-technical stakeholders in ensuring that the elicited NRs do not have well-formedness issues (WFIs)~\cite{Getir-Yaman-et-al-23,feng-et-al-23-b,feng-et-al-24,Getir-Yaman-et-al-23b} such as conflict and redundancy~\cite{Getir-Yaman-et-al-23b}.

\textbf{NR formalisation, validation, and limitations.} 
Previous work has proposed a domain specific language to formalize NRs~\cite{Getir-Yaman-et-al-23} as SLEEC rules. SLEEC is very close to natural language and has proven accessible to stakeholders without a technical background, including psychologist, philosophers, lawyers, ethicists, and regulators.
Automated techniques, based on first-order logic~\cite{feng-et-al-23} and a process algebra~\cite{Gibson-Robinson-et-al-14,DBLP:journals/acta/BaxterRC22}, have been developed to check normative requirements specified in this language for WFIs~\cite{Getir-Yaman-et-al-23,Getir-Yaman-et-al-23b,feng-et-al-23-b,feng-et-al-24}, and produces verification diagnosis accessible to non-technical stakeholders. While effective, these techniques assume that all capabilities 
referenced in the rules are independent. More precisely, the symbols used for describing events and measures in the rules are treated as uninterpreted constants that do not carry semantics. 
However, in practice, many of these capabilities are related, if we take into account implicit assumptions arising, for example, from domain knowledge or based on ordinary, everyday experiential knowledge~\cite{levesque-2017}.
Indeed, those relations are often implicitly assumed by non-technical stakeholders.
For example, `on' and `off' measures of a camera can be reasonably expected to be mutually exclusive. Ignoring these semantic relations can affect the soundness and completeness of the WFI analysis. For example, consider the SLEEC rules:

\smallskip
\noindent
\hspace*{5mm} $r_1$ = \sleec{\sleeckeyword{when} BatteryLow \sleeckeyword{then} MuteNotifications}\\
\hspace*{5mm}
$r_2$ = \sleec{\sleeckeyword{when} BatteryLow \sleeckeyword{then} SendUserWarning }

\smallskip
\noindent
In the rule set $\{r_1, r_2\}$, $r_1$ is conflicting with $r_2$. This is because $r_1$ and $r_2$ share the same triggering event \sleec{BatteryLow}, and the response of $r_1$, \sleec{MuteNotifications}, is contradictory with the response of $r_2$, \sleec{SendUserWarning}. The contradiction follows from the fact that, if the events \sleec{MuteNotifications} and  \sleec{SendUserWarning} occur simultaneously, that is, in the same time unit, they cannot both have their intended effect. When all goes well (that is, there is no failure), either a warning is delivered, or the agent is muted, but not both. However, existing automated reasoning techniques~\cite{Getir-Yaman-et-al-23,feng-et-al-23-b,feng-et-al-24} 
fail to detect this conflict as they do not capture the semantic relation between the two events used in the responses. Therefore, there is a need to define and capture semantic relations of terms used in NRs according to common sense and domain knowledge, and to use these relations in the NR analysis.

\textbf{Existing approaches to identify semantic relations.}
Traditional knowledge extraction approaches may be considered as a means for addressing this need. However, they use as input (structured or unstructured) sources without information specific to applications, while the semantic relations for capabilities used in SLEEC requirements come from common sense in the context of domain-specific environments.
Therefore, these knowledge extraction approaches are ineffective without an additional knowledge source, such as a domain-specific ontology~\cite{Farfeleder2005MKSO,Murugesh2015J,BencharquiOntologybasedRS, Diamantopoulos2018EnhancingRR} that is effort-intensive and costly to assemble. 
Our work addresses this issue without requiring such effort to structure the domain knowledge. 

\textbf{Our solution}.
We exploit the knowledge captured in Large Language Models (LLMs). LLMs are pre-trained with a vast amount of data covering a wide range of topics. They have been shown effective in answering questions regarding common sense~\cite{DBLP:conf/naacl/TalmorHLB19,DBLP:conf/aaai/BiskZLGC20,DBLP:journals/corr/abs-2303-11436} and domain-specific knowledge such as programming~\cite{DBLP:conf/aaai/JoshiSG0VR23,DBLP:conf/icer/PhungPCGKMSS22}, healthcare~\cite{DBLP:journals/jms/CascellaMBB23,haupt2023ai}, and law~\cite{DBLP:journals/corr/abs-2209-13020,DBLP:journals/corr/abs-2212-02199}. LLMs are also capable of obtaining new knowledge from analogies and examples~\cite{DBLP:conf/sigsoft/GuptaKBCGKRS023,DBLP:journals/corr/abs-2308-16118}. Therefore, we propose to use LLMs for suggesting potential semantic relations among system capabilities subject to SLEEC requirements. For the example above, GPT-4.0 successfully suggested that \sleec{MuteNotifications} is contradictory with \sleec{SendUserWarning}.
On the other hand, using LLMs comes with its own risks, as they are notorious for producing significantly incorrect results~\cite{DBLP:journals/corr/abs-2309-05922,DBLP:journals/corr/abs-2309-01219}. To prevent any falsely identified semantic relations from confounding WFI analysis, we also propose a lightweight logic-based filtering algorithm that incrementally extends relations via a set of inference rules and heuristically fixes logical inconsistencies on-the-fly. The filtered relations do not induce any logical fallacies and can be used directly for WFI analysis. We still, however, ask stakeholders to further filter the relations based on their domain-specific knowledge.

\input{Sections/fig-approach}

\textbf{Our contributions.} 
To summarise, the main contributions of our paper are:
(1)An LLM-based technique that extracts semantic relations between the capabilities of a system under development, where these capabilities appear as non-terminal symbols in the SLEEC rules that encode the system's NRs; 
(2) The integration of our new semantic-relation extraction technique with existing automated reasoning methods, to develop a systematic end-to-end approach, \approach, for eliciting and analyzing normative requirements; 
(3)  An extensive evaluation (with five non-technical stakeholders, using a range of real-world case studies) that shows the effectiveness and usability of our LLM-based technique and the \approach approach.

\textbf{Significance.}
This paper introduces an innovative approach to leverage LLMs to capture the implicit semantic relations that stakeholders inherently understand but often fail to articulate during the requirements elicitation process. By integrating LLMs, our technique not only facilitates bridging of the gap between the explicit formalization of NRs and the implicit understanding and assumptions of non-technical stakeholders, but also harmonizes this method with existing automated reasoning techniques. This integration allows for a more nuanced analysis of NRs, identifying potential conflicts and redundancies that may not be apparent without understanding the extracted underlying semantic relations.

The rest of the paper is organized as follows: ~Sec.~\ref{sec:background} gives the background material.
Sec.~\ref{sec:sanitization} presents \SAN, the new technique for determining semantic relations.   
Sec.~\ref{sec:approach} gives an overview of our approach, \approach, for eliciting and analyzing normative requirements. 
Sec.~\ref{sec:evaluation} presents the evaluation of the effectiveness of \SAN and \approach.
Sec.~\ref{sec:relatedwork} overviews related work.
Sec.~\ref{sec:conclusion} concludes the paper and discusses future research directions.

%% file: Sections/fig-approach.tex
\pdfoutput=1
\tikzset{
    stepop/.style={draw,dotted,rectangle,rounded corners,minimum width=1.5cm,minimum height=0.9cm,align=center,text width=3.3cm,fill=tuftsblue!30,font=\small\sffamily},
    stepopl/.style={draw,dotted,rectangle,rounded corners,minimum width=1.5cm,minimum height=0.9cm,align=center,text width=4cm,fill=tuftsblue!30,font=\small\sffamily},
    rndoutputitt/.style={draw,rectangle,rounded corners,minimum width=2cm,minimum height=1.5cm,align=center,text width=2.4cm,fill=tuftsblue!20,font=\small\sffamily},
    rndoutputi/.style={draw,rectangle,rounded corners,minimum width=2cm,minimum height=1.5cm,align=center,text width=3.2cm,fill=tuftsblue!20,font=\small\sffamily},
    rndoutputis/.style={draw,rectangle,rounded corners,minimum width=2cm,minimum height=0.9cm,align=center,text width=3.2cm,fill=tuftsblue!30,font=\small\sffamily},
    rndoutputib/.style={draw,rectangle,rounded corners,minimum width=4.2cm,minimum height=1.1cm,align=center,text width=4.2cm,fill=white!30,font=\small\sffamily},
    rndoutputibdona/.style={draw,rectangle,rounded corners,minimum width=3.2cm,minimum height=1.1cm,align=center,text width=3.2cm,fill=white!30,font=\small\sffamily},
    rndoutputibseva/.style={draw,rectangle,rounded corners,minimum width=2.6cm,minimum height=0.9cm,align=center,text width=2.6cm,fill=white!30,font=\small\sffamily},
    rndoutputiblina/.style={draw,rectangle,rounded corners,minimum width=3.5cm,minimum height=1.1cm,align=center,text width=3.2cm,fill=white!30,font=\small\sffamily},
    rndoutputibdo/.style={draw,dotted,rectangle,rounded corners,minimum width=2cm,minimum height=0.9cm,align=center,text width=3.5cm,fill=white!10,font=\small\sffamily},
    rndoutputibpu/.style={draw,rectangle,rounded corners,minimum width=2.7cm,minimum height=0.9cm,align=center,text width=3.8cm,fill=white!30,font=\small\sffamily},
    rndoutputibse/.style={draw,rectangle,rounded corners,minimum width=2cm,minimum height=0.9cm,align=center,text width=2.2cm,fill=tuftsblue!30,font=\small\sffamily},
    rndoutputibnl/.style={draw,rectangle,rounded corners,minimum width=2cm,minimum height=0.9cm,align=center,text width=4.2cm,fill=tuftsblue!30,font=\small\sffamily},
    rndoutputibp/.style={draw,rectangle,rounded corners,minimum width=2cm,minimum height=0.9cm,align=center,text width=2.45cm,fill=tuftsblue!30,font=\small\sffamily},
    rndoutputibc/.style={draw,line width=1pt,rectangle,rounded corners,minimum width=2cm,minimum height=0.9cm,align=center,text width=4.3cm,dotted,fill=gray!10,font=\small\sffamily},
    rndoutputibcs/.style={draw,line width=1pt,rectangle,rounded corners,minimum width=2cm,minimum height=0.9cm,align=center,text width=2.8cm,dotted,fill=gray!10,font=\small\sffamily},
    rndoutputl/.style={draw,dotted,rectangle,rounded corners,minimum width=0.45cm,minimum height=0.4cm,align=center,text width=0.75cm,fill=gray!30,font=\sffamily},
    rndoutputlg/.style={draw,rectangle,rounded corners,minimum width=0.45cm,minimum height=0.4cm,align=center,text width=0.75cm,line width=1pt,fill=gray!10,font=\sffamily},
    rndoutput/.style={draw,rectangle,rounded corners,minimum width=2cm,minimum height=1.5cm,align=center,text width=2cm,fill=tuftsblue!30,font=\small\sffamily},
    extact/.style={draw,dashed,rectangle,rounded corners,minimum width=2cm,minimum height=1.5cm,align=center,text width=3.4cm,dotted,fill=gray!50,font=\small\sffamily},
    recnorm/.style={draw,minimum width=0.6cm,minimum height=0.65cm,align=center,text width=0.65cm,dotted,fill=lightgray!50,font=\small\sffamily},
    recnormw/.style={draw,minimum width=0.6cm,dashed,minimum height=0.65cm,align=center,text width=0.65cm,fill=white,font=\sffamily},
    reclong/.style= {rounded corners=0.5cm,draw,minimum width=2cm,minimum height=1.5cm,align=center,text width=4cm,dotted,fill=gray!35,font=\sffamily},
    prog/.style={draw,dashed,tape,tape bend top=none,align=center,text width=2cm,fill=turquoisegreen!50,font=\small\sffamily},
    progl/.style={draw,tape,tape bend top=none,align=center,text width=0.7cm,fill=turquoisegreen!50,font=\small\sffamily},
    artifact/.style={draw,trapezium,trapezium left angle=75,trapezium right angle=105,text width=2.5cm,fill=cookie!35,align=center,font=\small\sffamily},
    artifactniki/.style={draw,trapezium,trapezium left angle=75,trapezium right angle=105,text width=1.5cm,fill=cookie!35,align=center,font=\small\sffamily},
    artifactnikita/.style={draw,trapezium,trapezium left angle=75,trapezium right angle=105,text width=1.57cm,fill=cookie!35,align=center,font=\small\sffamily},
    artifactm/.style={draw,trapezium,trapezium left angle=83,trapezium right angle=98,text width=2.7cm,fill=cookie!35,align=center,font=\small\sffamily},
    artifacti/.style={draw,trapezium,trapezium left angle=83,trapezium right angle=98,text width=2.3cm,fill=cookie!35,align=center,font=\small\sffamily},
    artifactitts/.style={draw,trapezium,trapezium left angle=83,trapezium right angle=98,text width=1.6cm,fill=cookie!50,align=center,font=\small\sffamily},
    artifactitt/.style={draw,trapezium,trapezium left angle=83,trapezium right angle=98,text width=2cm,fill=cookie!50,align=center,font=\small\sffamily},
    artifactittsss/.style={draw,trapezium,trapezium left angle=83,trapezium right angle=98,text width=1.7cm,fill=cookie!50,align=center,font=\small\sffamily},
    artifactittsssp/.style={draw,trapezium,trapezium left angle=83,trapezium right angle=98,text width=1.35cm,fill=cookie!50,align=center,font=\small\sffamily},
    artifactittlsi/.style={draw,trapezium,trapezium left angle=83,trapezium right angle=98,text width=2.55cm,fill=lightapricot!60,align=center,font=\small\sffamily},
    artifactittls/.style={draw,trapezium,trapezium left angle=83,trapezium right angle=98,text width=2.55cm,fill=cookie!50,align=center,font=\small\sffamily},
    artifactittlms/.style={draw,trapezium,trapezium left angle=83,trapezium right angle=98,text width=2.7cm,fill=cookie!50,align=center,font=\small\sffamily},
    artifactittl/.style={draw,trapezium,trapezium left angle=83,trapezium right angle=98,text width=3cm,fill=cookie!50,align=center,font=\small\sffamily},
    artifactittlsl/.style={draw,trapezium,trapezium left angle=83,trapezium right angle=98,text width=3.2cm,fill=cookie!50,align=center,font=\small\sffamily},
    artifactittll/.style={draw,trapezium,trapezium left angle=83,trapezium right angle=98,text width=3.6cm,fill=cookie!50,align=center,font=\small\sffamily},
    artifactittlll/.style={draw,trapezium,trapezium left angle=83,trapezium right angle=98,text width=3.8cm,fill=cookie!50,align=center,font=\small\sffamily},
    artifactittlls/.style={draw,trapezium,trapezium left angle=83,trapezium right angle=98,text width=3.15cm,fill=cookie!50,align=center,font=\small\sffamily},
    artifacto/.style={draw,trapezium,trapezium left angle=79,trapezium right angle=100,text width=1.5cm,fill=cookie!35,align=center,font=\small\sffamily},
    artifactoi/.style={draw,trapezium,trapezium left angle=79,trapezium right angle=100,text width=1.5cm,fill=lightgray!50,align=center,font=\small\sffamily},
    artifacts/.style={draw,dashed,trapezium,trapezium left angle=79,trapezium right angle=100,text width=3cm,fill=gray!50,align=center,font=\small\sffamily},
    artifactl/.style={draw,trapezium,minimum height=0.35cm,trapezium left angle=75,trapezium right angle=105,text width=0.3cm,fill=white,align=center,font=\sffamily},
    artifactvt/.style={draw,trapezium,trapezium left angle=83,trapezium right angle=98,text width=1cm,fill=cookie!35,align=center,font=\small\sffamily}
}

\begin{figure}[t]
    \centering
\scalebox{.88}{
\begin{tikzpicture}[x=2.25cm,y=0.9cm]

  \filldraw[draw=lightgray!60,rounded corners,fill=lightgray!10] (2.7,-5.5) rectangle (-1.78,1.82);
  \node[rotate=0,font=\large\sffamily] at (-1,1.4) {\textbf{\approach}};
\node[artifactittll] (preReqF) at (-0.6,-1.41)  {I.};
\draw[draw=lightapricot!80,rounded corners,fill=lightapricot!40] (-1.7,-1.1) rectangle  (0.29,-2.3);
\draw[draw=cookie!80,rounded corners,fill=cookie!20] (0.45,1.5) rectangle  (2.6,-2.55);
\node[rndoutputibseva] (conflict) at (1.06,-3.4)   {III.a. Conf};
\draw[draw=cottoncandy!60,rounded corners,fill=cottoncandy!60] (0.43,-2.7) rectangle  (2.6,-5);
\node[rndoutputibseva] (gost) at (-0.55,-3.4) {well};
\draw[draw=green!80,rounded corners,fill=green!20] (-1.6,-2.8) rectangle  (0.09,-3.9);
\node[artifactittl] (reqs) at (-0.6,-4.8) {well-formed \dsl requirements};

\node[rotate=0,font=\small\sffamily] at (-0.75,-1.6) {\textbf{I. Requirements elicitation}};
\node[rotate=0,font=\small\sffamily] at (1.5,0.9) {\textbf{II. Sanitizing definitions}};
\node[rotate=0,font=\small\sffamily] at (1.3,-3) {\textbf{III. WFI identification}};
\node[rotate=0,font=\sffamily] at (0.8,-3.6)  {$\square$ conflicts};
\node[rotate=0,font=\sffamily] at (2.05,-3.6)  {$\square$ insufficiency};
\node[rotate=0,font=\sffamily] at (1.0,-4.2)  {$\square$ restrictiveness};
\node[rotate=0,font=\sffamily] at (2.1,-4.2)  {$\square$ redundancy};
\node[inner sep=0pt] (autol) at (2.2,-3) {\includegraphics[scale=.025]{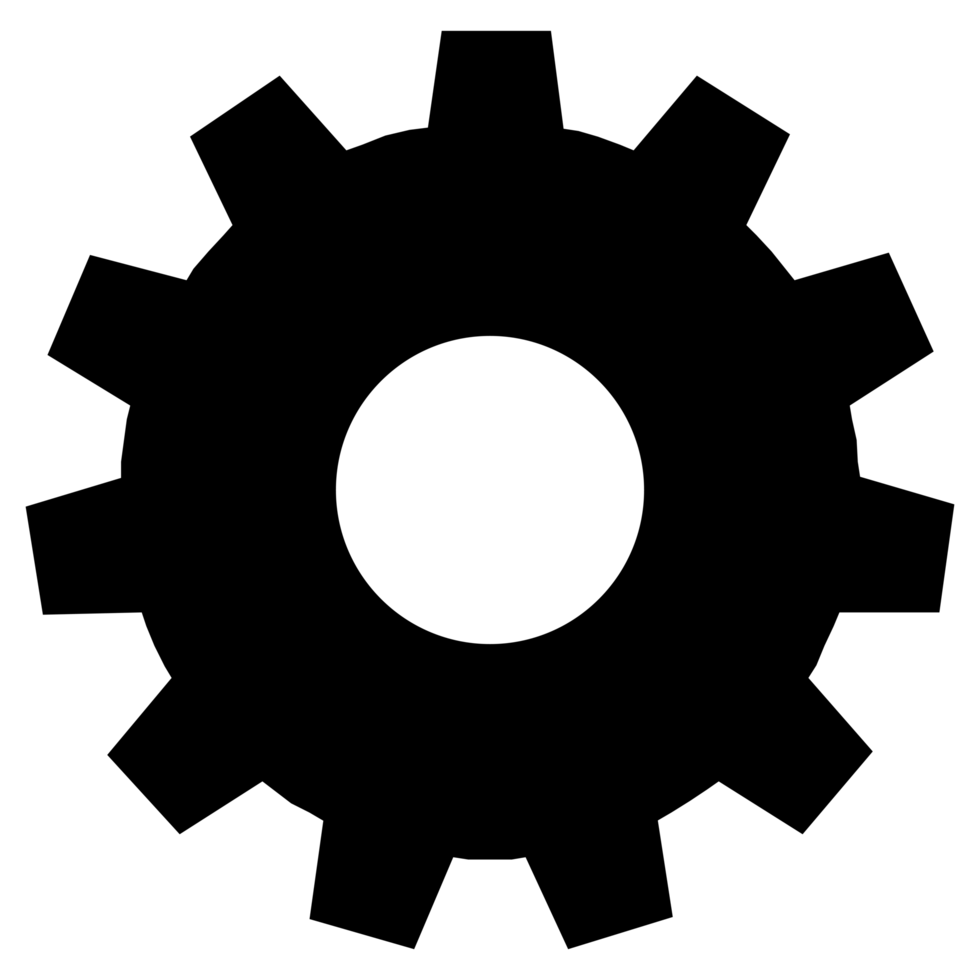}};
\node[rotate=0,font=\small\sffamily] at 
(-0.8,-3.15){\textbf{IV. WFI resolution}};
\node[inner sep=0pt] (manu) at (-0.5,-3.6) {\includegraphics[scale=.005,angle=55]{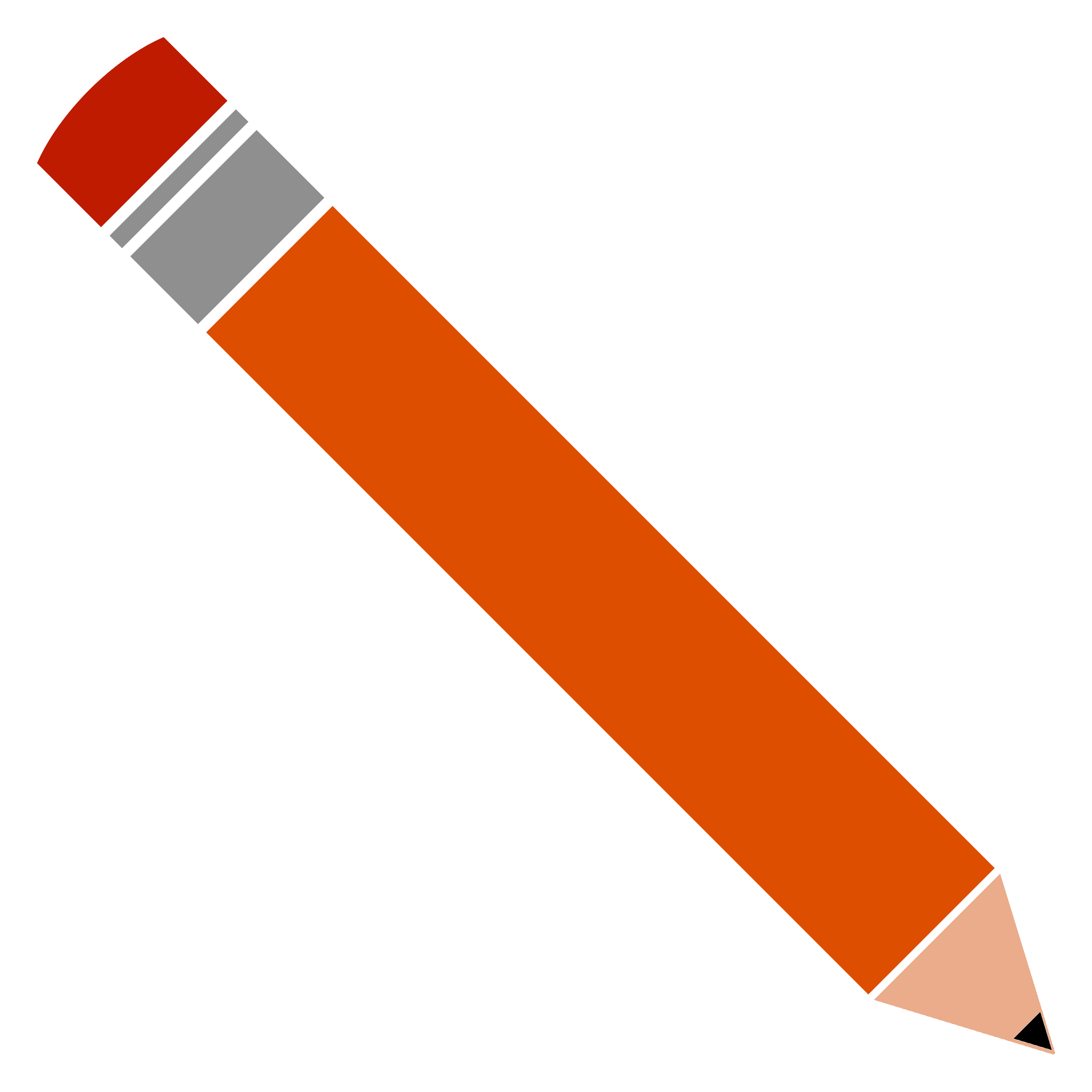}};
\node[inner sep=0pt] (autol) at (-1.35,-6.13) {\includegraphics[scale=.04]{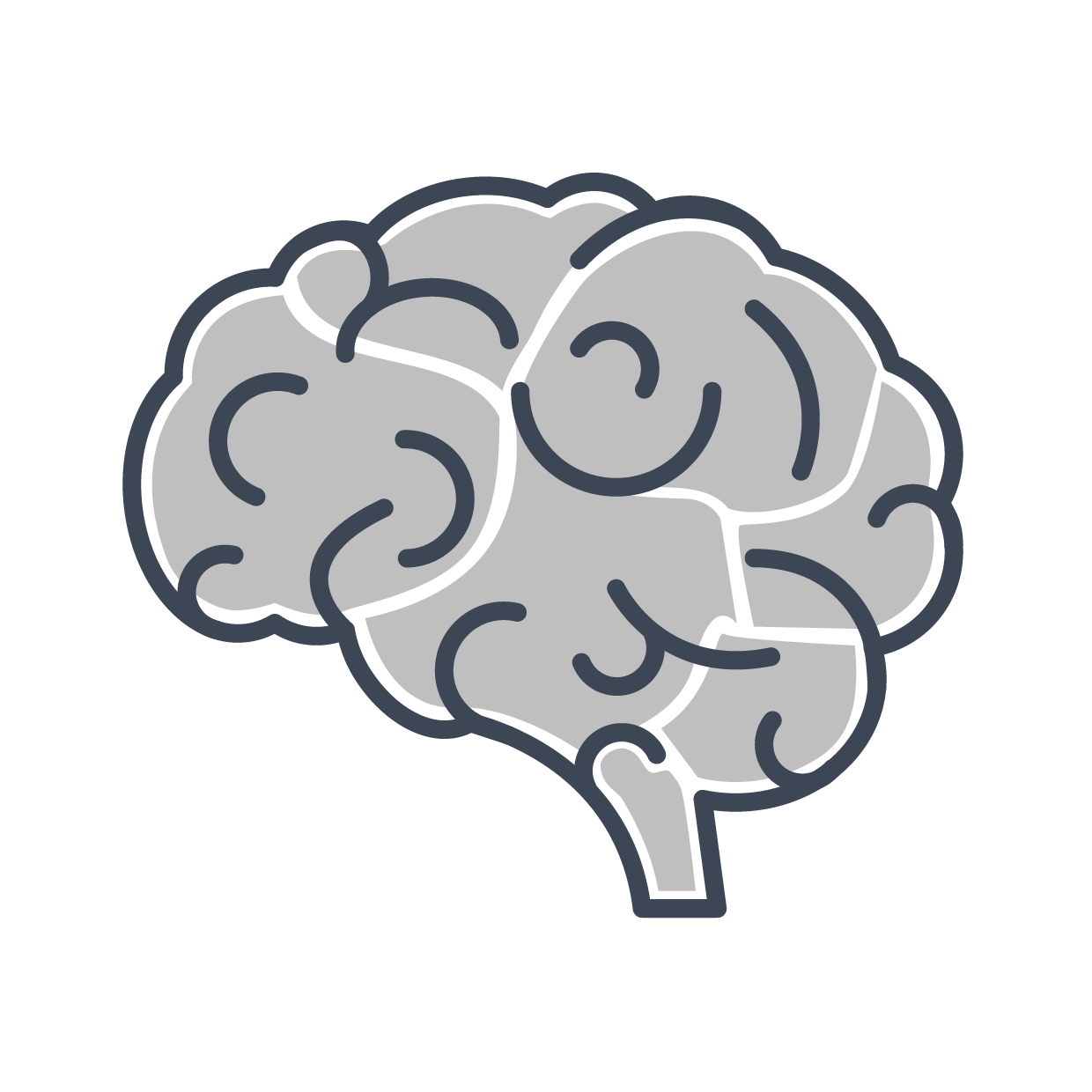}};
\node[rotate=0,font=\small\sffamily] at (-0.73,-6.13) {:~\SAN};
\node[inner sep=0pt] (autol) at (0.2,-6.13) {\includegraphics[scale=.035]{Sections/Img/auto1.png}};
\node[rotate=0,font=\small\sffamily] at (1.35,-6.13) {:~automated reasoning support};

\node[inner sep=0pt] (manu) at (-1.35,-7.1) {\includegraphics[scale=.005,angle=65]{Sections/Img/manu.png}};
\node[rotate=0,font=\small\sffamily] at (-0.7,-7.1) {:~Manual};

\node[artifactniki] (context) at (-1.2,0.1)  {SLEEC principles};
\node[artifactnikita] (capabilities) at (-0.2,0.1)  {system capabilities};
\node[inner sep=0pt] (manu) at (-0.55,-1.95) {\includegraphics[scale=.005,angle=55]{Sections/Img/manu.png}};
\node[rndoutputibpu] (sanitization) at (1.6,-0.2)   {II.a. Extracting\\ semantic relations~~};
\node[inner sep=0pt] (autol) at (2.37,-0.05) {\includegraphics[scale=.03]{Sections/Img/Logo-brain.png}};
\node[rndoutputibpu] (sanitizationResolve) at (1.6,-1.8)   {II.b. Review relations \\ ~};
\node[inner sep=0pt] (manu) at (1.65,-2.05) {\includegraphics[scale=.005,angle=55]{Sections/Img/manu.png}};

\begin{scope}[every path/.style={-latex},line width=1.2pt]
    \draw 
        (context) edge (preReqF)
        (capabilities) edge (preReqF)
        (preReqF) edge [bend left=5] (sanitization)
        (sanitization) edge (sanitizationResolve)
        (sanitizationResolve) edge (conflict)
        (conflict) edge (reqs)
        ;
    \draw[->,.style={-latex}] (gost) edge [bend left=10](conflict)
    (conflict) edge (gost);
\end{scope}
\end{tikzpicture}
}
\caption{\small Overview of \approach: the no\underline{R}m\underline{A}tive requ\underline{I}reme\underline{N}ts eli\underline{C}itati\underline{O}n and v\underline{A}lida\underline{T}ion approach. 
}
    \label{fig:approach}
    \vspace{-0.1in}
\end{figure}

%% file: Sections/background.tex
\section{Preliminaries}
\label{sec:background}

In this section, we present background material. In Sec.~\ref{ssec:ndsl}, we review the details of SLEEC, and in Sec.~\ref{ssec:wfi}, we present well-formedness issues of SLEEC rules.

\begin{table}[t]
    \centering
    \caption{\small Syntax of normalized $\dsl$ with signature ($\events$, $\measures$).}
    \scalebox{0.8}{
        \begin{tabular}{ll}
        \toprule
        Name & Definition\\
        \hline
        Term & $\term$ := $\constant :\mathbb{N} \mid \measure\in \measures \mid -\term \mid \term + \term \mid \constant \times \term$  \\
        Proposition & $\prop$ := $\top \mid \bot \mid \term = \term \mid \term \ge \term \mid \notS \; \prop \mid \prop \; \andS \; \prop \mid \prop \; \orS \; \prop$ \\
        Obligation & $\obg^+$ := $ \within{\event}{\term}$ $ \mid$ $\obg^-$ := $\within{\notS \; \event}{\term}$ \\
        Cond Obligation & $\cobg^+$ := $\prop \Rightarrow \obg^+$ $\mid$ $\cobg^-$ := $\prop \Rightarrow \obg^-$ \\
        Obligation Chain & $\dobg $ := $\cobg \mid \cobg^+ \otherwise \dobg$ \\
        Rule & $\srule $ := $\rulesyntax{\event \sand \prop }{\dobg}$ \\
        \hline
        Fact & $\fact$ := $\factsyntax{\event \sand \prop}{(\dobg \mid \notS \dobg)}$\\
        \bottomrule
        \\
        \end{tabular}
    }
 \label{tab:syntax}
 \vspace{-0.18in}
\end{table}

\subsection{\dsl}\label{ssec:ndsl}
 \dsl is an event-based language for specifying normative requirements through the pattern ``\textbf{when} trigger \textbf{then} response'' ~\cite{Getir-Yaman-et-al-23}. It builds on propositional logic enriched with temporal constraints (e.g., \textbf{within} x \textbf{minutes}) and the constructs \textbf{unless}, specifying \emph{defeaters}, and \textbf{otherwise}, specifying \emph{fallbacks}.
 
In more detail, a \dsl rule has the basic form: \sleec{\sleeckeyword{when} Trigger \sleeckeyword{then} Response}. The trigger defines an event whose occurrence leads to the need to satisfy the constraints defined by the response. 
For example, Rule10 in Tab.~\ref{tab:reqs} applies when an event \sleec{MeetingUser} occurs. In addition, a trigger may include a condition~(Boolean expression) over measures. 
For instance, if the trigger of Rule10 is changed to \sleec{MeetingUser \sleeckeyword{and} \sleeckeyword{not} patientStressed}, then triggering of Rule10 additionally requires the value of the boolean measure \sleec{patientStressed} to be false when the event \sleec{MeetingUser} occurs. 

The response defines events that need to occur when the rule is triggered, and may include deadlines and timeouts. For example, the response in RuleA defined as  \sleecT{\sleeckeyword{when} MeetingPatient \sleeckeyword{then} ExaminingPatient \sleeckeyword{within} 30 \sleeckeyword{minutes}} specifies an obligation for the event \sleecT{ExaminingPatient} to occur within 30 minutes since the triggering of RuleA. To accommodate situations where a response may not happen within its deadline, the \sleecT{\sleeckeyword{otherwise}} construct can be used to specify an alternative response. For example, if the response of RuleA is changed to \sleecT{ExaminingPatient \sleeckeyword{within} 30 \sleeckeyword{minutes} \sleeckeyword{otherwise} VisitingPatient \sleeckeyword{within} 5 \sleeckeyword{minutes}}, then if the primary response \sleecT{ExaminingPatient} times-out after 30 minutes, then the alternative 
requires that \sleecT{VisitingPatient} occurs within 5 minutes after the timeout.

\dsl allows a rule to include one or more defeaters, introduced
by the \sleeckeyword{unless} keyword, and specifying circumstances that preempt the original response and provide an alternative. For example, in a RuleB defined as \sleecT{\sleeckeyword{when} MeetingPatient \sleeckeyword{then} ExaminingPatient \sleeckeyword{with\-in} 5 \sleeckeyword{minutes} \sleeckeyword{unless} patientStressed \sleeckeyword{then} \sleeckeyword{not} Examining\-Patient \sleeckeyword{with\-in} 2 \sleeckeyword{minutes}}, the original response \sleecT{ExaminingPatient} is preempted if \sleecT{patientStressed} is true when RuleB is triggered. Instead, the defeater defines an alternative response that forbids any occurrence of the event  \sleecT{ExaminingPatient}  within 2 minutes of triggering of RuleB.  

\dsl is an expressive language, allowing specification of complex nested defeaters and responses.
A \dsl rule may be represented by multiple logically equivalent formulations, complicating analysis and reasoning.  This issue can be addressed with the \textit{normalized} \dsl, which simplifies nested response frameworks and removes defeaters, encoding their semantics using propositional logic.
A translation from \dsl to the normalized form is provided in~\cite{feng-et-al-24-e}.  The syntax of the normalized \dsl is given in Tab.~\ref{tab:syntax} and the  semantics is described in
Appendix~\ref{app:semantics}.

\begin{table}[t]
    \centering
    \caption{{\small Capabilities, definitions and requirements of  DAISY~\cite{daisy-22}.}}
    \label{tab:mainTable}
    \vspace{-0.02in}
    \begin{minipage}{\linewidth}
        \centering
        \begin{subtable}{\linewidth}
            \centering  
            \caption{{\small DAISY capabilities.}}
            \label{tab:capabilities}
            \scalebox{0.72}{
                \begin{tabular}{l l l}
                \toprule
                - Meet a patient & - Issue commands to the patient & - Explain the protocols to the patient \\
                - Examine the patient & - Collect patient data & - Perform triage assessment \\
                - Provide instructions & - Repeat the instructions  & - Request clinician intervention \\
                \bottomrule
                \\
                \end{tabular}
            }
        \end{subtable}
        
        \vspace{0.1em}

        \begin{subtable}{\linewidth}
            \centering
            \caption{{\small Sample DAISY \dsl definitions.}}
            \label{tab:defs}
            \scalebox{0.9}{
                \begin{tabular}{cl | cl}
                \toprule
                \sleeckeyword{event} & 
                \sleec{ExamineState} & \sleeckeyword{measure} & \sleec{patientStressed} \\
                \sleeckeyword{event} & \sleec{IdentifyDAISYTrust} & \sleeckeyword{measure} & \sleec{patientAgeConsidered}  \\
                \sleeckeyword{event} & \sleec{MeetingUser}  & \sleeckeyword{measure} & \sleec{patientXReligion} \\
                \sleeckeyword{event} & \sleec{MeetingPatient}  & \sleeckeyword{measure} & \sleec{userDirectsOtherwise} \\
                \bottomrule
                \\
                \end{tabular}
            }   
            \vspace{0.1em} 

    \begin{subtable}{\linewidth}
        \centering
        \caption{{\small Sample DAISY \dsl requirements.}}
        \label{tab:reqs}
        \scalebox{0.9}{
            \begin{tabular}{cl}
            \toprule
        	\sleec{Rule10} & \sleec{\sleeckeyword{when} MeetingUser \sleeckeyword{then} ExamineState} \\
        	\sleec{Rule13} & \sleec{\sleeckeyword{when} MeetingUser \sleeckeyword{then} IdentifyDAISYTrust} \\
            \sleec{Rule16} & \sleec{\sleeckeyword{when} MeetingUser \sleeckeyword{then} ExaminingPatient} \\
            & \sleec{\sleeckeyword{within} 30 \sleeckeyword{minutes} \sleeckeyword{unless} patientStressed} \\ 
            \sleec{Rule19} & \sleec{\sleeckeyword{when} MeetingUser \sleeckeyword{and} patientXReligion}\\
            & \sleec{\sleeckeyword{then not} ExaminingPatient} 
        	   \\
        	& \sleec{\sleeckeyword{unless} userDirectsOtherwise} \sleec{\sleeckeyword{unless} medicalEmergency} \\ 
            \bottomrule
            \\
            \end{tabular}
        }
    \end{subtable}
            
        \end{subtable}
    \end{minipage}
\end{table}
\input{Sections/figTraces}

\subsection{Well-Formedness Issues (\WFI)}
\label{ssec:wfi}
\dsl rules $\ruleset$ might be subject to well-formedness issues (WFIs): \emph{vacuous conflict},  \emph{redundancy}, \emph{insufficiency}, \emph{over-restrictiveness} and \emph{conditional conflict}. We  explain each WFI briefly and refer to~\cite{feng-et-al-24} for formal definitions.

\boldparagraph{Vacuous conflict} A rule is  \emph{vacuously conflicting in a rule set $\ruleset$} if triggering this rule always leads to a conflict with some other rules in 
$\ruleset$.
\begin{example}\label{example:vc}
    Let $\ruleset$ consist of the following rules: 
\begin{itemize}
    \item \sleec{R1}: \sleecT{\sleeckeyword{when} A \sleeckeyword{then} B \sleeckeyword{within} 5 \sleeckeyword{hours}}
    \item \sleec{R2}: \sleecT{\sleeckeyword{when} B \sleeckeyword{then} C \sleeckeyword{within} 5 \sleeckeyword{hours}}
    \item \sleec{R3}: \sleecT{\sleeckeyword{when} A \sleeckeyword{then} \sleeckeyword{not} C \sleeckeyword{within} 10 \sleeckeyword{hours}}
\end{itemize}
\sleec{R1} is vacuously conflicting in  this set because of \sleec{R2} and \sleec{R3}. Similarly, \sleec{R3} is vacuously conflicting because of \sleec{R1} and \sleec{R2}. 
\end{example}

\boldparagraph{Situational conflict} A rule is \emph{situational conflicting in a rule set $\ruleset$} if triggering this rule under a certain situation, which defines a history of events and measures, always leads to a conflict with some other rules in the future.

\begin{figure}
    \centering
    \begin{tabular}{c}
        \scalebox{.75}{
            \begin{tikzpicture}[scale=2]
              \node at (1.9,0.9) {\small\color{mygray}time};
              
              \node at (0.5,1.24) {\small\color{mymauve}A};
              \node at (1.5,1.24) {\small\color{mymauve}B};
              
              \node at (0.4,0.92) {\small\color{mygray}$1$};
              \node at (1.5,0.92) {\small\color{mygray}$6$};
              
              \node at (-0.5,1.0) {$\sigma$};
        
              \draw [thick,black,->] (-0.3,1) -- (2.2,1);
              \draw [thick,black,-] (-0.3,1.05) -- (-0.3,0.95);
              \draw [thick,black,->] (0.4,1) -- (0.4,1.15);
              \draw [thick,black,->] (1.5,1) -- (1.5,1.15);
              
            \end{tikzpicture}
        }
    \end{tabular}
    \caption{{A conflicting situation of \sleec{R2} in Example~\ref{example:sc}.}}
    \label{fig:sctrace}
\end{figure}
\begin{example}\label{example:sc}
    Let a  rule set $\ruleset$ with rules \sleec{R2} and \sleec{R3} from Example~\ref{example:vc} be given. \sleec{R2} is situational conflicting in $\ruleset$  with the situation $\sigma$ shown in  Fig.~\ref{fig:sctrace}. In $\sigma$, \sleec{R3} is triggered at hour 1, which forbids any occurrence of event C until hour 11. However, \sleec{R2} is triggered at hour 2, and this causes a conflict because C must occur between hours 5 and 11.
\end{example}

\begin{figure}
    \centering \shadowbox{\includegraphics[width=0.47\textwidth,trim={0cm 0cm 0.8cm 0.1cm}, clip]{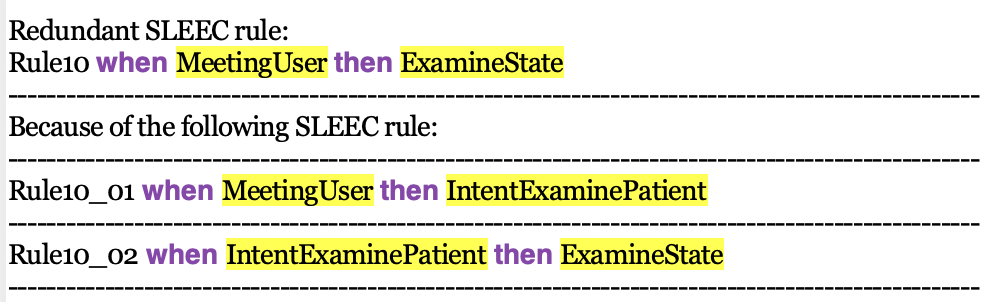}}
    \caption{{\small Redundancy diagnosis example.}}
    \label{fig:redundantDiag}
    \vspace{-0.2in}
\end{figure}

\boldparagraph{Redundancy} A rule is  \emph{redundant in a rule set $\ruleset$} if this rule is logically implied by other rules in $\ruleset$.
\begin{example}\label{example:red}
    In the example  in Fig.~\ref{fig:redundantDiag},  \sleec{R10} is redundant because \sleec{R10} is logically implied by the combination of \sleec{R10\_01} and \sleec{R10\_02}.
\end{example}

\boldparagraph{Insufficiency} A rule set $\ruleset$ is \emph{insufficient subject to a given concern} if some behavior represented by the concern is realizable while respecting $\ruleset$.

\boldparagraph{Over-restrictiveness} A rule set $\ruleset$ is \emph{overly restrictive subject to a given purpose} if none of the functional goals represented by the purpose is realizable while respecting $\ruleset$.

\smallskip

 WFIs impose threats to the validity of \dsl rules. Conflicting rules must be resolved to avoid inevitable SLEEC harms. Situational conflicts and redundancies also require attention. The former capture dead-end situations where conflicts become inevitable and the latter complicates  management of  evolving requirements. Insufficient  and  overly-restrictive rule sets also need to be fixed, to avoid executions of undesirable behaviors and violations of functional goals, respectively.

\boldparagraph{Order of Resolving WFIs}
Different types of WFIs have different levels of severity and might depend on each other. 

Vacuous conflicts should be resolved prior to situational conflicts because any situation that triggers a vacuously conflicting rule will also create a conflicting situation. For example, consider \sleecT{R1 = \sleeckeyword{when} A \sleeckeyword{then} B}. This rule is vacuously conflicting due to another rule: \sleecT{R2 = \sleeckeyword{when} A \sleeckeyword{then} \sleeckeyword{not} B \sleeckeyword{within} 10 minutes}. This implies that \sleec{R1} is also situationally conflicting since any situation that triggers \sleec{R1} leads to a conflict. 
Insufficiency and over-restrictiveness are checked and resolved after both types of conflict because the behaviors defined by conflicting rules are inconsistent, and thus might affect the judgment of insufficiency or over-restrictiveness analysis. For example, without first resolving the vacuous or situational conflict between \sleecT{R1} and \sleecT{R2}, the conflicting situations are excluded from the insufficiency and over-restrictiveness analysis. The concern \sleecT{\sleeckeyword{exists} A \sleeckeyword{while} X} can never be identified in any currently conflicting situation (and cannot be identified at all if a vacuous conflict exists), thus it is blocked by the conflict.

 Redundancies are checked last because they are the least severe WFIs, and because they are affected by conflicts.  For example, without resolving the vacuous conflict between \sleecT{R1} and \sleecT{R2}, the rule \sleec{R3 = \sleeckeyword{when} A \sleeckeyword{then} C} is redundant because triggering \sleec{R3} would also trigger \sleec{R1} and subsequently cause a conflict. Moreover, the conflict subsumes the response of \sleec{R3} (e.g., \sleecT{C}), thus making the rule redundant.
 
{\tool}~\cite{feng-et-al-24}  identifies and resolve WFIs. It encodes the semantics of rules into \fol and turns WFI conditions into \fol satisfiability queries. The satisfiability results, either satisfying solutions or proofs of unsatisfiability, are then used to diagnose the causes of WFIs. 
\begin{example}
    Consider a rule set $\ruleset$ $= \{$\sleec{Rule10}, \sleec{Rule10\_01}, \sleec{Rule10\_02}$\}$, where \sleec{Rule10,Rule10\_01,Rule10\_02} are rules in Tbl.~\ref{tab:reqs}.
    To check whether \sleec{Rule10} is redundant in $\ruleset$, we can use \tool which checks the satisfiability of the query $\{\neg$\sleec{Rule10} $\} \cup \{r \mid r \in$ \sleec{Rule10}$\setminus \{$\sleec{Rule10}$\}\}$. The query is unsatisfiable (UNSAT), which means that \sleec{Rule10} is a logical consequence of the $\{\neg$\sleec{Rule10}$\} \cup \{r \mid r \in$ \sleec{Rule10}; hence, \sleec{Rule10} is redundant. \tool provides a diagnosis shown in Fig.~\ref{fig:redundantDiag}, with the reason of redundancy highlighted.
\end{example}

%% file: Sections/figTraces.tex
\begin{figure}
    \centering
    \begin{tabular}{c}
        \scalebox{.75}{
            \begin{tikzpicture}[scale=2]
              \node at (1.9,0.9) {\small\color{mygray}time};
              
              \node at (0.5,1.24) {\small\color{mymauve}ExaminingPatient};
              \node at (0.5,1.44) {\small\color{mymauve}patientStressed};
              \node at (1.5,1.24) {\small\color{mymauve}$\emptyset$};
              
              \node at (0.4,0.92) {\small\color{mygray}$1$};
              \node at (1.5,0.92) {\small\color{mygray}$30$};
              
              \node at (-0.5,1.0) {$\sigma_1$};
        
              \draw [thick,black,->] (-0.3,1) -- (2.2,1);
              \draw [thick,black,-] (-0.3,1.05) -- (-0.3,0.95);
              \draw [thick,black,->] (0.4,1) -- (0.4,1.15);
              \draw [thick,black,->] (1.5,1) -- (1.5,1.15);
              
            \end{tikzpicture}
        }
    \end{tabular}
    \caption{{\small Trace for the DAISY robot example.}}
    \label{fig:trace}
\end{figure}

%% file: Sections/sanitization.tex
\section{Sanitizing Definitions}
\label{sec:sanitization}


The semantics of \dsl (see Sec.~\ref{ssec:ndsl}) assumes that the capabilities of the systems (represented by events and measures) are independent. \dsl does not support the description of the relations between these capabilities. In this section, we first propose an extension to \dsl for describing the relations between capabilities, denoted as CR (Sec.~\ref{ssec:CR}). Then, we present an LLM-aided approach to automatically extract CRs from the textual semantics of the capabilities (Sec.~\ref{ssec:LLMCR}).

\subsection{Capability Relations~(CR)}
\label{ssec:CR}

We propose extending \dsl to capture three types of binary relations: (1) between two events, (2) between two measures, and (3) between an event and a measure. 

\boldparagraph{Relations between events}
We capture the following list of binary relations between events $\event_a$ and $\event_b$:
\begin{itemize}
\item $\event_{a}$ \sleeckeyword{hypernym} $\event_{b}$ iff (if and only if) for every state, the occurrence of $\event_{a}$ implies the occurrence of $\event_{b}$.
\item $\event_{a}$ \sleeckeyword{isContradictoryWith} $\event_{b}$ iff for every state, $\event_{a}$ and $\event_{b}$ never occur simultaneously.
\item $\event_{a}$ \sleeckeyword{happensBefore} $\event_{b}$ iff for every state $s_1$ where $\event_{a}$ occurs, there exists some state $s_2$ prior to $s_1$ in which $\event_{b}$ occurred.
\item $\event_{a}$ \sleeckeyword{equal} $\event_{b}$ iff $\event_{a}$ \sleeckeyword{hypernym} $\event_{b}$ and $\event_{b}$ \sleeckeyword{hypernym} $\event_{a}$.
\end{itemize}
The formal definitions of these relations are presented in the extended version.
Table~\ref{tab:CReventsExample} illustrates  each relations.

\boldparagraph{Relations between measures}
We capture the following list of binary relations between propositions $\prop_a$ and $\prop_b$ over measures:
\begin{itemize}
\item $\prop_a$ \sleeckeyword{imply} $\prop_b$ iff for every state, $\prop_a$ holding implies that $\prop_b$ also holds.
\item $\prop_a$ \sleeckeyword{mutuallyExclusive} $\prop_b$ iff for every state, $\prop_a$ and $\prop_b$ never hold simultaneously.
\item $\prop_a$ \sleeckeyword{opposite} $\prop_b$ iff for every state, $\prop_a$ holds if and only if $\prop_b$ does not hold.
\item $\prop_a$ \sleeckeyword{equal} $\prop_b$ iff $\prop_a$ \sleeckeyword{imply} $\prop_b$ and $\prop_b$ \sleeckeyword{imply} $\prop_a$.
\end{itemize}
Formal definitions of these relations are in the supplementary material~\cite{RE-Artifact}. 
Tbl.~\ref{tab:CRmeasureExample} illustrates  each relation type. 

\boldparagraph{Relations between events and measures}
We capture the following list of relations between events and measures:

\begin{itemize}
\item $\prop$ \sleeckeyword{forbids} $\event$ iff
 for every state, if $\prop$ holds, it implies that the occurrence of $\event$ is not possible at those times.
\item $\event$ \sleeckeyword{induces} $\prop$ iff for every state, the occurrence of $\event$ implies that $\prop$ holds at the same timeg.
\item \sleeckeyword{when} $\event_a$ \sleeckeyword{then} $\prop$ \sleeckeyword{until} $\event_b$ iff $\prop$ holds for all states starting from a state where $\event_a$ occurs until a state where $\event_b$ occurs.
\item \sleeckeyword{when} $\event$ \sleeckeyword{then} $\prop$ \sleeckeyword{for} $\term$ iff $\prop$ holds for all states starting from a state $s_1$ where $\event$ occurs until a state $s_2$ such that $s_2$ is $\term$ units of time after $s_1$.
\end{itemize}
The formal definitions of the above relations are presented in Appendix~\ref{app:relsem}.
Table~\ref{tab:EMrelationsExample} gives examples of each type of relation. Note that \sleeckeyword{forbids} and \sleeckeyword{induces} describe relations that are concerned with the same state (i.e., with the same time point), while \sleeckeyword{when} $\ldots$ \sleeckeyword{then} $\ldots$ \sleeckeyword{until} $\ldots$ and \sleeckeyword{when} $\ldots$ \sleeckeyword{then} $\ldots$ \sleeckeyword{for} $\ldots$ are temporal relations between events and measures across multiple states. 

\boldparagraph{Rationale behind the selection of semantic relation terms}
The terms used to describe the semantic relations between events and measures were determined through collaboration with a non-technical stakeholder, a philosopher with expertise in common-sense knowledge and intuition-based reasoning. The relations were chosen to be easy to understand and intuitive where possible. The majority of the terms have everyday usage across multiple contexts and domains, thereby making them easier for stakeholders to adopt. These ‘common sense’ relations include `happens before’, `equal’, `imply’, `opposite’, `forbids’, `contradictory' and `induces’. Stakeholders will likely already possess an understanding of these relations prior to being introduced to them in the context of events and measures (although each relation is still thoroughly explained to stakeholders and illustrated with examples to aid comprehension). 
%
In the absence of readily apparent intuitive or common-sense equivalents and to avoid verbose descriptions, terms were selected that instead capture the intended semantic relation. `Hypernym' was selected based on its frequent use in semantic relation databases such as WordNet~\cite{Miller-90,Lin-et-al-20,Snow-et-al-04}. The term `mutually exclusive’ was chosen due to its frequent use in philosophy, law, computer science and other relevant disciplines. To ensure concurrence across disciplines, the precise meaning intended for this work was specified and standardised (as was the case for each relation identified).

\begin{table}[t]
    \caption{{\small Example relations between events.}}
    \centering
    \scalebox{0.8}{
        \begin{tabular}{c|rcl}
        \toprule
             Relation & \multicolumn{3}{c}{SLEEC DSL example} \\
        \midrule
             \multirow{2}{*}{hypernym} & \multicolumn{3}{l}{\textit{water is an instance of liquid}}  \\
             & \multirow{1}{*}{\sleec{DrinkWater}} & \multirow{1}{*}{\sleeckeyword{hypernym}} & \multirow{1}{*}{\sleec{DrinkLiquid}}\\
             \multirow{2}{*}{equal} & \multicolumn{3}{l}{patient and client are interchangeable, when patients are the only clients} \\
             & \sleec{CallPatient} & \sleeckeyword{equal} & \sleec{CallClients}\\
             \multirow{2}{*}{contradictory} & \multicolumn{3}{l}{\textit{impossible to open and close the door simultaneously}}  \\
              & \multirow{1}{*}{\sleec{OpeningDoor}} & \multirow{1}{*}{\sleeckeyword{isContradictoryWith}} & \multirow{1}{*}{\sleec{ClosingDoor}} \\
             %
             happen & \multicolumn{3}{l}{\textit{locking the door occurs after closing it}} \\
             before & \multirow{1}{*}{\sleec{ClosingDoor}} & \multirow{1}{*}{\sleeckeyword{happensBefore}}& \multirow{1}{*}{\sleec{LockingDoor}}\\
        \bottomrule
        \end{tabular}
    }
    \label{tab:CReventsExample}
\end{table}


\begin{table}[t]
    \caption{{\small Example semantic relations between measures.}}
    \centering
    \scalebox{0.8}{
        \begin{tabular}{rl}
        \toprule
             Notation & \multicolumn{1}{c}{SLEEC DSL example}\\
        \midrule
             \multirow{2}{*}{imply} & \multicolumn{1}{l}{\textit{opened door can not be locked}} \\
             & \sleec{doorOpened} \sleeckeyword{imply} \sleec{not doorLocked}\\
             \cellcolor{almond!40}  mutual & \cellcolor{almond!40} \textit{opened door  can not be locked} \\
             \cellcolor{almond!40} exclusive & \cellcolor{almond!40}  \sleec{doorOpened} \sleeckeyword{mutuallyExclusive} \sleec{doorLocked}\\
              \multirow{2}{*}{opposite} & \multicolumn{1}{l}{\textit{the door can be either opened or closed}} \\
              & \sleec{doorOpened} \sleeckeyword{oppositeTo} \sleec{doorClosed} \\
              \cellcolor{almond!40} equal & \cellcolor{almond!40} \textit{patient and user are interchangeable, when patients are the only users} \\
              \cellcolor{almond!40} & \cellcolor{almond!40} \sleec{patientConsented} \sleeckeyword{equal} \sleec{userConsented}\\
        \bottomrule
        \end{tabular}
    }
    \label{tab:CRmeasureExample}
\end{table}


\begin{table}[t]
    \caption{{\small Example dependencies between events and measures.}}
    \centering
    \scalebox{0.8}{
        \begin{tabular}{rl}
        \toprule
             ID & \multicolumn{1}{c}{SLEEC DSL example}\\
        \midrule
        \cellcolor{almond!40} induce & \cellcolor{almond!40} \sleec{CollectConsent \sleeckeyword{induces} consentObtained}\\
         forbidden & \sleec{inWater \sleeckeyword{forbids} CarStartSpeeding}\\
        \cellcolor{almond!40} until & \cellcolor{almond!40} \sleec{\sleeckeyword{when} CollectConsent \sleeckeyword{then} consentObtained} \\
        \cellcolor{almond!40} & \cellcolor{almond!40} \sleec{\sleeckeyword{until} ConsentWithdraw}\\
             for & \sleec{\sleeckeyword{when} LoginConfirmed \sleeckeyword{then} loggedIn \sleeckeyword{for} 10 \sleeckeyword{minutes}}\\
        \bottomrule
        \end{tabular}
    }
    \label{tab:EMrelationsExample}
\end{table}

\subsection{LLM-aided CR Extractions}
\label{ssec:LLMCR}

\begin{algorithm}[t]
\caption{{\small \SAN($\events$, $\measures$)}}\label{alg:extractAlg}
\begin{algorithmic}[1]
\Require $\events$ and $\measures$ are the complete sets of event and measure symbols, respectively
\Ensure $Rel_f$ is a consistent set of relations between events and measures in $\events$ and $\measures$, respectively.
\State $query \gets \textsc{PrepareQuery(\events, \measures})$\label{ln:preparequery}
\State $Rel \gets \textsc{LLM}(query)$\label{ln:executequery}
\State $Rel_f \gets \textbf{Filter}
(Rel)$ \label{ln:filter}
\State \Return $Rel_f$
\end{algorithmic}
\end{algorithm}
\begin{algorithm}[t]
\caption{{\small \textbf{CheckConsistency}($Rel$)}}\label{alg:consistencyChecking}
\begin{algorithmic}[1]
\Require $Rel$ is a set of binary relations between two events or two measures.
\Ensure $Rel_f$ is a subset of $Rel$, and is logically consistent w.r.t. the inference rules in Fig.~\ref{fig:inferrules}.
\State $Rel* \gets Rel$, $Rel' \gets \emptyset$ 
\label{ln:initRel}
\While{$\top$}
    \State $Rel* \gets Rel* \bigcup Rel'$\label{ln:integrate}
    \State $Rel* \gets \{r \mid 
 r \text{ is negative } \vee \negate{r} \not\in Rel* \}$\label{ln:localcon}
    \State $old\_Rel* \gets Rel*$
    \State $Rel* \gets \textsc{ApplyRules}^-(Rel*)$\label{ln:applynegative}
    \State $Rel*$, $Rel? \gets \textsc{ApplyRules}^+(Rel*)$\label{ln:applypositive}
\If{$old\_Rel* = Rel*$ $\wedge$ $Rel? = \emptyset$}\label{ln:checkfixpoint}
    \State \Return $Rel \bigcap Rel*$ \label{ln:return}
\EndIf

    \State $Rel' \gets \textsc{LLM}(Rel?)$\label{ln:followup}
\EndWhile
\end{algorithmic}
\end{algorithm}
\begin{figure*}
    \centering
    \footnotesize
    $
    \begin{gathered}
    \toprule
    \textbf{Inference rules for relations between two events.}\\
    \toprule
    \infer[(IP1^-)]{\negate{\hyp{\event_a}{\event_b}}}{%
        \con{\event_a}{\event_b}
    }
    \;\;
    \infer[(IP2^-)]{\negate{ \hyp{\event_a}{\event_b}}}{%
        \hb{\event_a}{\event_b}
    } 
    \;\;
    \infer[(IPtrans^+)]{ \hyp{\event_a}{\event_c}}{%
        \hyp{\event_a}{\event_b} \land  \hyp{\event_b}{\event_c} 
    } 
    \;\;
    \infer[(IPEQ^+)]{ \hyp{\event_a}{\event_b}}{%
        \eq{\event_a}{\event_b} 
    } 
    \\
    \;\;
    \infer[(ME1^-)]{ \negate{ \con{\event_a}{\event_b}}}{%
        \hyp{\event_a}{\event_b} 
    } 
    \;\;
    \infer[(MEcomm^+)]{ \con{\event_a}{\event_b}}{%
        \con{\event_b}{\event_a}
    } \;\;
    \infer[(MEtrans^+)]{ \con{\event_a}{\event_c}}{%
        \hyp{\event_a}{\event_b} \land \con{\event_b}{\event_c} 
    } 
    \\
     \infer[(EQIP^+)]{ \eq{\event_a}{\event_b}}{%
        \hyp{\event_a}{\event_b} \land 
        \hyp{\event_b}{\event_a} 
    }
    \;\;
    \infer[(EQcom^+)]{ \eq{\event_b}{\event_a}}{%
        \eq{\event_a}{\event_b} 
    }\\
    \infer[(HBtrans1^+)]{ \hb{\event_a}{\event_c}}{%
        \hb{\event_a}{\event_b} \land \hb{\event_b}{\event_c}
    }\;\;
    \infer[(HBtrans2^+)]{ \hb{\event_a}{\event_c}}{%
        \hb{\event_a}{\event_b} \land \hyp{\event_c}{\event_b}
    }\\
    \infer[(HBtrans3^+)]{ \hb{\event_a}{\event_c}}{%
        \hyp{\event_b}{\event_a} \land \hb{\event_b}{\event_c}
    }\\
    \end{gathered}
    $

    $
    \begin{gathered}
    \\
    \toprule
    \textbf{Inference rules for relations between two propostions over measures.}\\
    \toprule
    \infer[(MIP1^-)]{\negate{\imply{\prop_a}{\prop_b}}}{%
        \me{\prop_a}{\prop_b}
    }
    \;\;
    \infer[(MIPtrans^+)]{ \imply{\prop_a}{\prop_c}}{%
        \imply{\prop_a}{\prop_b} \land  \imply{\prop_b}{\event_c} 
    } 
    \;\;
    \infer[(IPEQ1^+)]{ \imply{\prop_a}{\prop_b}}{%
        \eq{\prop_a}{\prop_b} 
    } \;\;
    \infer[(IPEQ2^+)]{ \imply{\neg \prop_a}{\neg \prop_b}}{%
        \eq{\prop_a}{\prop_b} 
    }
    \\
    \;\;
    \infer[(MME1^+)]{ \negate{ \me{\prop_a}{\prop_b}}}{%
        \imply{\prop_a}{\prop_b} 
    } 
    \;\;
    \infer[(MMEcomm^+)]{ \me{\prop_a}{\prop_b}}{%
        \me{\prop_b}{\prop_a}
    } \;\;
    \infer[(MMEtrans^+)]{ \me{\prop_a}{\prop_c}}{%
        \imply{\prop_a}{\prop_b} \land \me{\prop_b}{\prop_c} 
    } \\
     \infer[(MEQIP^+)]{ \eq{\prop_a}{\prop_b}}{%
        \imply{\prop_a}{\prop_b} \land 
        \imply{\prop_b}{\prop_a} 
    }
    \;\;
    \infer[(MEQOP^+)]{ \eq{\prop_a}{\prop_b}}{%
        \ops{\prop_a}{\neg \prop_b} \land 
        \imply{\prop_b}{\prop_a} 
    }
    \;\;
    \infer[(MEQcom^+)]{ \eq{\prop_b}{\prop_a}}{%
        \eq{\prop_a}{\prop_b} 
    }\\
    \infer[(MOPEQ^+)]{ \ops{\prop_a}{\prop_c}}{%
        \eq{\prop_a}{\neg \prop_b}
    }\;\;
    \infer[(MOPcoms^+)]{ \ops{\prop_a}{\prop_b}}{%
        \ops{\prop_b}{\prop_s} 
    }\;\;
    \infer[(MOPME^+)]{ \ops{\prop_a}{\prop_b}}{%
        \me{\prop_a}{\prop_b} \land \me{\neg \prop_a}{\neg \prop_b}
    }\\
    \bottomrule
    \end{gathered}
    $
    
    \caption{{\small Horn inference rules for binary relations between two events or two measures.  Rules with  trailing  \quoted{+}s and \quoted{-}s derive positive and negative relations, respectively. Alg.~\ref{alg:consistencyChecking} always propagates on negative (\quoted{-}) before propagating on positive (\quoted{+}) rules.} 
    }
    \label{fig:inferrules}
    \vspace{-0.1in}
\end{figure*}

Given sets of symbols $\events$ and $\measures$ for the events and measures of a SLEEC document, the algorithm \SAN (see  Alg.~\ref{alg:extractAlg}) computes potential binary relations between the event and measure symbols in $\events$ and $\measures$ according to their textual semantics. \SAN first prepares and executes queries (Alg.~\ref{alg:extractAlg} LL:\ref{ln:preparequery}-\ref{ln:executequery})
to an LLM to extract a set of candidate relations (as presented in Sec.\ref{sssec:llminit}). Then, it filters (L:\ref{ln:filter}) the candidate set to ensure logical consistency (as presented in Sec.~\ref{sssec:filter}).

\begin{remark}
   \SAN automatically extracts relations only between  a pair of events or measures.  Relations between an event (and multiple events, in the case of \sleeckeyword{when} \sleeckeyword{until}) and a measure should be more carefully defined by the users. 
\end{remark}

\subsubsection{Relation discovery}\label{sssec:llminit}
\begin{figure}[t]
\includegraphics[width=0.38\textwidth,trim={0cm 0.8cm 0cm 0.8cm}, clip]{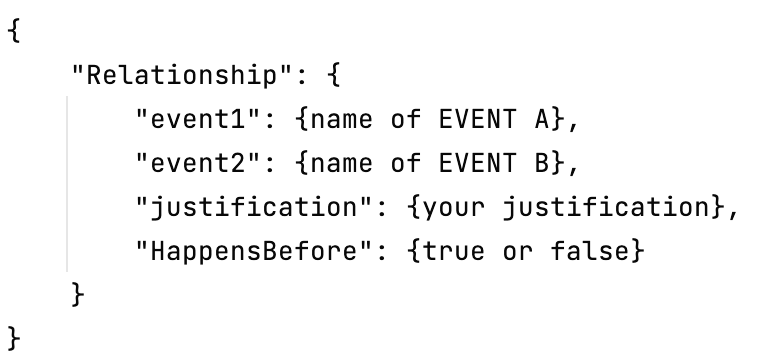}
    \caption{{\small Output format for  binary relations of \quoted{\sleeckeyword{happensBefore}}.}}
    \label{fig:figexpected}
\end{figure}

\SAN initially queries an LLM to discover a set of binary relations between two events or two measures. The query to the LLM includes the following information: (1)
the grammar of \dsl (shown in Table~\ref{tab:syntax});
(2) the set of event and measure symbols in $\events$ and $\measures$, respectively;
(3) the textual definition and an illustration for each type of relation. For example, the query for the relation \sleeckeyword{happensBefore} is shown in Fig.~\ref{fig:defexample};
(4) the JSON format for the expected output. For example, the output format for a discovered \sleeckeyword{happensBefore} relation is shown in Fig.~\ref{fig:figexpected}.

\begin{figure}[t]
\fbox{
    \begin{minipage}{\dimexpr0.5\textwidth-1cm}
        For every pair of events $\event_a$ and $\event_b$, can you please answer the following question: Is it the case that there is always an occurrence of $\event_a$ that happens before every occurrence of $\event_b$? If yes, please say $\hb{\event_a}{\event_b}$. If no, please say $\negate{\hb{\event_a}{\event_b}}$. For example, \sleecT{\hb{CreateForm}{ShowForm}} is correct because \sleecT{CreateForm} is the prerequisite of \sleecT{ShowForm}.
    \end{minipage}
    }
    \caption{{\small LLM  query for finding binary relations of \quoted{\sleeckeyword{happensBefore}}.}}
    \label{fig:defexample}
\end{figure}

\begin{figure}[t]
\includegraphics[width=0.5\textwidth,trim={0cm 0.8cm 0cm 0.8cm}, clip]{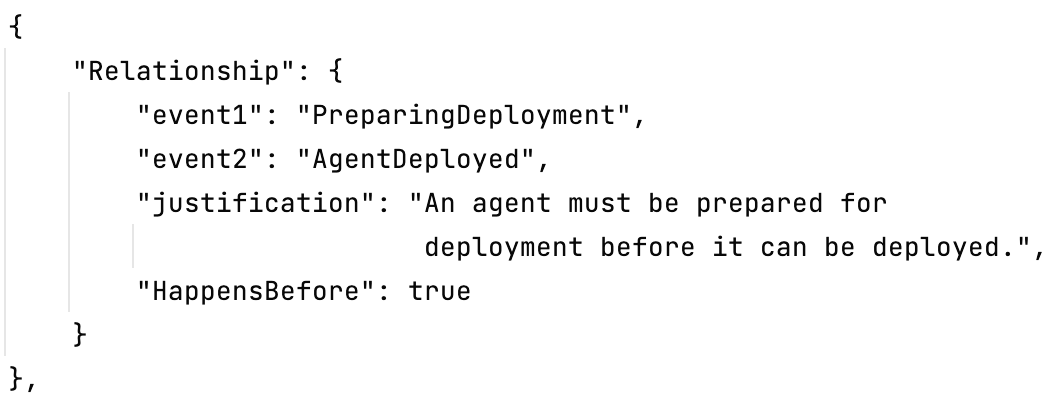}
\vspace{-0.2in}
    \caption{{\small LLM output for  a positive relation of  \quoted{\sleeckeyword{happensBefore}.}}}
    \label{fig:llmoutputexample}
\end{figure}

\SAN sends a query to the LLM and then waits for it to populate a JSON file containing a set of discovered relations and their verdicts $Rels ={r_1, \ldots, r_n}$. A relation is considered positive if it is confirmed by the LLM (e.g., $\hb{\event_a}{\event_b}$) and negative if it is rejected (e.g., $\negate{\hb{\event_a}{\event_b}}$). Fig.~\ref{fig:llmoutputexample} demonstrates the LLM's output representing a positive relation of \sleeckeyword{happensBefore}. For all relations, \SAN asks the LLM to provide a justification before producing the final verdict. This simulates the forward-thinking process and seems to prevent the LLM from retrofitting a justification for a random verdict. In our experiments, we have observed that if LLMs are tasked with the same or similar queries multiple times and prompted to provide justification before the final verdict, they are less likely to produce conflicting results and illogical justifications.

\boldparagraph{Prompt Engineering}
    To produce a prompt for querying the LLM, we have first experimented with a number of different queries and incrementally refined them to optimize output consistency: the LLM should not provide conflicting results for the same or similar queries. The work has been guided by a collection of small examples, and the final result reported here evaluated on a collection of real examples as discussed in Sec.~\ref{sec:evaluation}.  In the initial experiments, we asked the LLM to associate a confidence score with each relation verdict, but the results showed that the provided confidence scores were highly artificial and non-deterministic.
    Moreover, the initial experiments also indicated that the LLM tends to over-generalize the semantics of relations and wrongly detect more of them than we had expected. Inspired by the idea of learning by demonstrations, we have added examples (e.g., the last sentence in Fig.~\ref{fig:defexample}) to the prompt to illustrate each relation.

\subsubsection{Filtering relations via logic inference}\label{sssec:filter}
The initial set of candidate relations $Rel$ produced by the LLM might contain semantic errors that could lead to logical inconsistencies and subsequently confound the WFIs analysis. While \SAN, as an automated procedure, cannot find all semantic errors in $Rel$, it addresses any potential logical inconsistencies in it.

\SAN calls Alg.~\ref{alg:consistencyChecking} on $Rel$ to remove candidate relations that could lead to logical inconsistencies, based on the inference rules shown in Fig.~\ref{fig:inferrules}. These 
inference rules derive new relations from existing ones in $Rel$.  Let $Rel*$ be the fixed-point set of all relations that are either in $Rel$ or derivable from $Rel*$. A relation $r$ \textit{witnesses} an inconsistency if $Rel*$ contains both the positive relation $r$ and the negative relation $\negate{r}$. A relation set $Rel$ is \textit{consistent} if $Rel*$ has no inconsistency witnesses. 
Alg.~\ref{alg:consistencyChecking} takes as argument a set $Rel$ of relations and returns a candidate set $Rel_{f} \subseteq Rel$ that is consistent.

All inference rules in Fig.~\ref{fig:inferrules} are Horn clauses in the form $r_1, r_2, \ldots, r_n \Rightarrow r_{cons} \mid \negate{r_{cons}}$. Only positive relations in $Rel$ can lead to the deduction of new relations using the rules. Negative relations are used for identifying inconsistency witnesses. Therefore, an empty set of relations is \emph{vacuously consistent}. It is, however,  not very useful. \SAN aims to greedily find a reasonably large subset of $Rel$ that is consistent. It is \textit{not guaranteed}, however, that \SAN  returns the largest consistent subset, as the problem of finding this  subset is generally NP-hard, even for Horn logic~\cite{DBLP:journals/ipl/JaumardS87}.

Given an input set $Rel$, Alg.~\ref{alg:consistencyChecking} iteratively computes $Rel*$ from $Rel$ (L:\ref{ln:initRel}) while fixing any identified inconsistency on-the-fly. For each iteration, before applying inference rules, \SAN first checks $Rel*$ for local consistency (L:~\ref{ln:localcon}) and overrides all 
inconsistent positive relations $r \in Rel*$ (i.e., if $\negate{r} \in Rel*$). Negative relations are favored over positive ones to speed-up convergence since all inference rules are Horn. After fixing local consistencies,
all rules for deriving negative relations (i.e., inference rules whose name ends with \quoted{-}) are exhaustively applied (L:\ref{ln:applynegative}). When an inference rule derives a negative relation $\negate{r_{cons}}$, there are three cases: (1) $\negate{r_{cons}} \in Rel*$, (2) $r_{cons} \in Rel*$, and (3) otherwise. In the first case, the consequence is consistent with $Rel*$ and no action is required. In the second case, there is an inconsistency in $Rel*$ witnessed by $r_{cons}$. Alg.~\ref{alg:consistencyChecking} then resolves the inconsistency by updating $r_{cons}$ with $\negate{r_{cons}}$ in $Rel*$. In the third case, $\negate{r_{cons}}$ is added to $Rel*$.

\begin{example}
    Let $Rel = \{r1 = \hyp{\event_a}{\event_b},  \; r2= \hb{\event_a}{\event_b}\}$ be given.  Alg.~\ref{alg:consistencyChecking} first derives a relation $r3 = \negate {\hyp {\event_a}{\event_b}}$ from  $r2$ by rule \quoted{$IP2^-$}. The derived relation $r3$ conflicts with $r_1$ in $Rel$ (and $Rel*$). The conflict is resolved by replacing $r_1$ with $r_3$ in $Rel*$. 
\end{example}

After exhausting all negative inference rules and obtaining a new $Rel*$ (L:\ref{ln:applynegative}), Alg.~\ref{alg:consistencyChecking} then proceeds to 
{
apply the positive rules (L:\ref{ln:applypositive}). If an inference rule derives a positive relation $r_{cons}$, there are three cases: (1) $r_{cons} \in Rel*$, (2) $\negate{r_{cons}} \in Rel*$, and (3) otherwise. In the first case, the consequence is consistent. In the second case, there is an inconsistency in $Rel*$ witnessed by $r_{cons}$. Alg.\ref{alg:consistencyChecking} resolves this inconsistency by updating one of the premises $r_i \in r_1, \ldots, r_n$ with $\neg r_i$ in $Rel*$. In the third case, instead of directly adding $r_{cons}$ to $Rel*$, Alg.~\ref{alg:consistencyChecking} prepares a follow-up query to the LLM about $r_{cons}$ for confirmation. Note that this follow-up query is necessary because the LLM did not initially discover $r_{cons}$ as a positive relation in $Rel$. This might imply that $\negate{r_{cons}} \in Rel$ if the LLM follows the closed-world assumption, or `unknown' otherwise.
We store all follow-up queries 
to the LLM in a set $Rel?$ when applying the positive inference rules. An example of the follow-up query is shown in Fig.~\ref{fig:additionalquery}.

After exhausting the positive inference rules, the follow-up queries $Rel?$ are sent to the LLM (L:\ref{ln:followup}). The results, denoted as $Rel'$, comprise a set of positive and negative relations, representing confirmations and rejections of the queries in $Rel?$, respectively. Alg.~\ref{alg:consistencyChecking} then merges $Rel*$ with $Rel'$ (L:\ref{ln:integrate}), and starts a new iteration of inferences (LL:\ref{ln:applynegative}-\ref{ln:followup}). Alg.~\ref{alg:consistencyChecking} terminates when $Rel*$ reaches a fixed point and no follow-up queries are produced (L:\ref{ln:checkfixpoint}). Finally, Alg.~\ref{alg:consistencyChecking} returns the intersection of $Rel*$ and the original relations $Rel$ (L:\ref{ln:return}) as the consistent set of relations. Note that not every relation in $Rel*$ can be derived from the final return set because some 
relations  might be updated after being used to derive other relations in $Rel*$. We do not  include these non-derivable relations in the return even though including them does not affect consistency.
On the other hand, 
the derivable relations  in $Rel*$
do not need to be in the final return set since they can be logically inferred, and thus are redundant.

\begin{figure}[t]
    \centering
\includegraphics[width=0.5\textwidth]{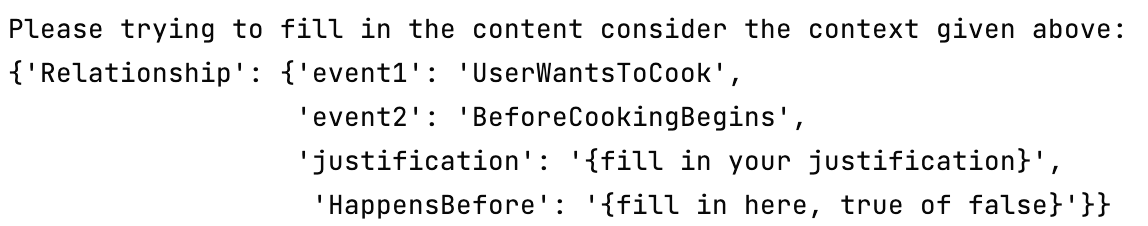}
    \caption{Follow-up query for confirming the relation \sleec{\hb{UserWantsToCook}{BeforeCookingBegins}}.}
    \label{fig:additionalquery}
\end{figure}
\begin{example}
    Let $Rel = \{r1 = \hyp{\event_a}{\event_b},  \; r2= \con{\event_b}{\event_c}, r3 = \hyp{\event_a}{\event_c}\}$  be given. Running Alg.~\ref{alg:consistencyChecking} first derives a positive  relation $r4 = \me {\event_a}{\event_c}$ from $r1$ and $r2$ by rule \quoted{$METrans^+$}. $r4$ and $\negate{r4}$ are not in $Rel*$, and $r4$ is added to $Rel?$ to be queried by the LLM. Suppose the LLM confirms $r4$ and adds it to $Rel*$. Then in the next iteration, Alg.~\ref{alg:consistencyChecking} derives $r5 = \negate{\hyp{\event_a}{\event_c}}$ from $r4$ by rule \quoted{$IP1^-$}, which conflicts with $r1$ in $Rel$. The conflict is resolved by replacing $r_1$ with $r5$ in $Rel*$.  At this point, $Rel* = \{r2, r3, r4, r5\}$ is a consistent set of relations. Alg.~\ref{alg:consistencyChecking} returns the intersection of $Rel$ and $Rel*$, which is $\{r2, r3\}$.
\end{example}

\begin{remark}
    Alg.~\ref{alg:consistencyChecking}  terminates. This is because (1) there is a finite number of possible binary relations, and (2) every relation $r$ in $Rel*$ can change at most once: a positive relation $r$ can be updated to $\neg r$, but $\neg r$ can never be updated to $r$.
\end{remark}
\noindent
\SAN returns the consistent relation set $Rel_f$ 
as the candidate relations. Stakeholders are expected to review these candidates and validate the correct ones, which are then automatically included in the \dsl rules for WFIs analysis.






%% file: Sections/approach.tex
\section{The \approach Approach}
\label{sec:approach}

In this section, we introduce our tool-supported approach, \approach, for no\underline{r}m\underline{a}tive requ\underline{i}reme\underline{n}ts eli\underline{c}itati\underline{o}n and v\underline{a}lida\underline{t}ion, depicted  
in Fig.~\ref{fig:approach}.
The approach consists of four distinct stages which we discuss below.


\boldparagraph{I. Initial requirements elicitation}
The goal of this stage is to systematically identify \textit{preliminary normative requirements} and obtain their \textit{formal} and \textit{machine-readable} representations. 

\approach follows the approach presented in \cite{townsend-et-al-2022} to first contextualize the high-level SLEEC principles 
by mapping them onto the agent capabilities.
Consider DAISY, the robotics system described in Sec.~\ref{sec:background}.  Its
 capabilities to \quoted{meet a patient} and \quoted{explain the interaction protocols to the patient} are mapped to the SLEEC principles \quoted{autonomy} 
and \quoted{self-determination}, and the mapping could result in the following preliminary requirement:  ``when DAISY meets a patient, it shall explain the interaction protocol to her''. 
Note that during the elicitation process, new capabilities might be recommended by non-technical stakeholders if they believe that a SLEEC harm could occur otherwise. For instance, to comply with privacy legislation, stakeholders could recommend adding the \quoted{RemoveData} capability to DAISY, which grants users the right to be forgotten.

For formalizing preliminary requirements, \approach follows the approach presented in \cite{feng-et-al-24}, which uses the domain-specific language \dsl. 
This DSL has been shown to be accessible to stakeholders from different fields, including lawyers, philosophers, ethicists, roboticists, and software engineers~\cite{Getir-Yaman-et-al-23,Getir-Yaman-et-al-23b,feng-et-al-23-b,feng-et-al-24}. 
For example, the preliminary autonomy requirement is formalized as a \dsl rule   \sleecT{\sleeckeyword{when} MeetingPatient \sleeckeyword{then} ExplainProtocols}. 
More examples of DAISY's \dsl requirements and their definitions are provided in Tbl.~\ref{tab:reqs} and \ref{tab:defs}, respectively.

\boldparagraph{II. Sanitizing definitions}
This stage aims to enrich the preliminary set of normative requirements by capturing and integrating the semantic relations (see Sec.~\ref{ssec:CR}) between system capabilities (i.e., events and measures). This stage uses our LLM-aided CR extraction algorithm \SAN (see Sec.~\ref{ssec:LLMCR}) to compute a set of candidate relations which are then (manually) validated by the stakeholders. The validated relations are integrated into the preliminary normative requirements. For example, the  relation 
\sleecT{MeetingUser \sleeckeyword{equal} MeetingPatient} 
has been suggested, validated and integrated for DAISY because patients are the only DAISY users.  


%

\begin{figure}
    \centering
    \shadowbox{\includegraphics[width=0.47\textwidth,trim={0cm 0cm 0.8cm 0.6cm}, clip]{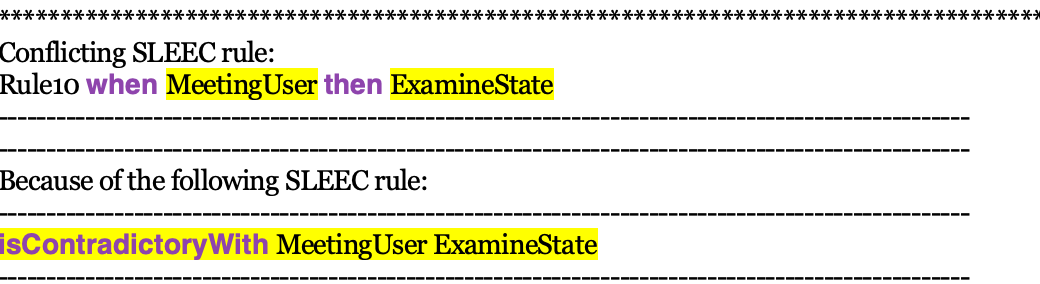}}
    \caption{{\small A diagnosis showing that \sleecT{Rule10} and the relation \sleecT{MeetingUser \sleeckeyword{isContradictoryWith} with ExamineState} cause a vacuous conflict. }}
    \label{fig:dFigure}
    \vspace{-0.2in}
\end{figure}

\boldparagraph{III. Identification of well-formedness issues}
In this stage, we detect WFIs (see Sec.\ref{ssec:wfi}) in the preliminary normative requirements using the existing automated reasoning technique, \tool, in the 
order
(1) \emph{vacuous conflicts}, (2) \emph{situational conflicts}, (3) \emph{insufficiencies} \& \emph{over-restrictiveness}, and (4) \emph{redundancies}.
For each issue, we provide a \emph{diagnosis} to help stakeholders understand exactly which rules and clauses caused the \WFI. 
Consider a rule set containing \sleecT{Rule10} and a relation \sleecT{MeetingUser \sleeckeyword{isContradictoryWith} with ExamineState}.  A vacuous conflict is identified and a diagnosis (see Fig.~\ref{fig:dFigure}) is produced to show that \sleecT{Rule10} and the relation are the causes of the conflict. 

If an WFI is identified, \approach jumps to the (manual) `WFI Resolution' stage (Stage IV).
The process terminates when no further WFIs have been identified, resulting in well-formed SLEEC requirements.

\boldparagraph{IV. WFI Resolution}
The objective in this stage is for the user to use the provided diagnosis to (manualy) address the  identified \WFI{s}.
For example, consider
resolving the vacuous conflict given the diagnosis in Fig.~\ref{fig:dFigure}.
We can deduce that the rule should not require a simultaneous response. Instead, it could be rewritten to include a time delay, such as: \sleecT{\sleeckeyword{when} MeetingUser \sleeckeyword{then} ExamineState \sleeckeyword{within} 10 \sleeckeyword{minutes}}.
Stakeholders are expected to resolve each issue with the exception of redundancies, where they may choose to intentionally preserve some of them.  
After resolving the WFIs, the stakeholders return to Stage~III to confirm that the refinement process did not  introduce new WFIs.
If the resolution process involved updating some of the requirements definitions, the stakeholders should return to Stage II to capture the potential dependencies in the updated capabilities.

%

\approach 
actively engages stakeholders in the elicitation and rule analysis process, and completes when the stakeholders have resolved all conflicts and concerns, and deem  the elicited requirements to be comprehensive. Our experiments (see Sec.~\ref{sec:evaluation}) show that the process successfully terminates
in 1-3 iterations.





%% file: Sections/evaluation.tex
\vspace{-0.1in}
\section{Evaluation}
\label{sec:evaluation}

\subsection{Evaluation methodology}
We carried out an extensive two-part study to answer the following research questions (RQs).

\smallskip
\noindent
\textbf{RQ1:} Does identifying semantic relations with \SAN improve the effectiveness of analyzing normative requirements WFIs as measured by the  number of relevant and spurious issues identified?
By answering this question, we study the effectiveness of identifying semantic relations (i.e., domain engineering) in improving  requirements analysis.

\smallskip
\noindent
\textbf{RQ2:} How well does \approach support non-technical stakeholders in guiding the elicitation and analysis process?
By answering this question, we aim to evaluate the effectiveness of \approach w.r.t. (a) its ability to facilitate the elicitation of a comprehensive set of requirements that encompass all existing system capabilities as well as social, legal, ethical, empathetic, and cultural considerations; and (b) its impact on supporting non-technical stakeholders in achieving a well-formed (e.g., conflict-free) set of
requirements.

\smallskip
\noindent
\textbf{Common activities carried out for the two RQs.}
Given real-world cases, non-technical stakeholders (N-TSs)
were asked to review the capabilities relation inferred by \SAN on the normative requirements they had elicited for these cases. The premise was that those who elicited the requirements should also validate the correctness of the identified semantic relations. 
For each case, the semantic relationships inferred by \SAN were presented to the N-TSs using a Google Form. In the form, N-TSs individually labeled these semantic relationships as correct or incorrect, provided justifications if necessary, and timed the process. 
After identifying the relations classified as correct, we analyzed the cases for any new well-formedness issues. We then asked the stakeholders to review these newly identified \WFI{s} to ensure they were \emph{relevant}, i.e., not spurious.

\smallskip
\noindent
\textbf{Reproducibility.} To enable the reproducibility of our results, we have made \SAN implementation and all our evaluation artifacts available in~\cite{RE-Artifact}.

\begin{table*}[t] 
    \caption{\small \SAN effectiveness. We record: true positive (TP) and false negative (FP) in the extracted capabilities relation (rel.); the number of hyponyms (hyp.), coincidences (coi.), contradicting (cont.), happening before (h.b.), implications (imp.), mutually exclusive (m.e.), equivalences (eq.), induce (induc), and forbid (forb.) relations added; the number of new issues identified (after capturing the semantic relations) for vacuous conflict (v-conf.), situational conflict (s-conf.), redundancy (redund.), restrictiveness (restrict.), and insufficiency (insuff.).}
    \label{tab:SAN-eval}
    \centering
    \scalebox{0.8}{
        \begin{tabular}{c c || c | c c c c c c c c c | c c c c c}
            \toprule
            \multirow{2}{*}{cases} & 
            rules & relations & \multicolumn{9}{c}{number of relations by type}  & \multicolumn{5}{c}{new \WFI} \\
            & (evnt.$-$ msr.) & (TP - FP) & \multirow{1}{*}{hyp.} & \multirow{1}{*}{coi.} & \multirow{1}{*}{cont.} & \multirow{1}{*}{h.b.} & \multirow{1}{*}{imp.} & \multirow{1}{*}{m.e.} & \multirow{1}{*}{eq.} & \multirow{1}{*}{induc.} & \multirow{1}{*}{forb.} & \multirow{1}{*}{v-conf.} & \multirow{1}{*}{s-conf.} & \multirow{1}{*}{redund.} & \multirow{1}{*}{insuffi.} & \multirow{1}{*}{restrict.}\\ \toprule
            ALMI & $39$ ($41-15$) & 5 - 2 & 0 & 0 & 0 & 0 & 2 & 2 & 1 & 0 & 0 & 0 & 0 & 0 & 0 & 0 \\
            \cellcolor{gray!10} ASPEN & \cellcolor{gray!10} $15$ ($25-18$)& \cellcolor{gray!10} 19 - 8 & \cellcolor{gray!10} 0 & \cellcolor{gray!10} 0 & \cellcolor{gray!10} 7 & \cellcolor{gray!10} 1 & \cellcolor{gray!10} 2  & \cellcolor{gray!10} 7 & \cellcolor{gray!10} 1 & 
            \cellcolor{gray!10} 0 & 
            \cellcolor{gray!10} 1 & \cellcolor{gray!10} 1 & \cellcolor{gray!10} 0 & \cellcolor{gray!10} 1 & \cellcolor{gray!10} 0 & \cellcolor{gray!10} 0 \\
            AutoCar & 19 (36-26) & 6 - 19 & 0 & 0 & 5 & 0 & 0 & 0 & 0 & 0 & 1 & 0 & 0 & 0 & 0 & 0 \\
            \cellcolor{gray!10} BSN & \cellcolor{gray!10} 29 ($33$-$31$)& \cellcolor{gray!10} 1 - 3 & \cellcolor{gray!10} 0 & \cellcolor{gray!10} 0 & \cellcolor{gray!10} 1 & \cellcolor{gray!10} 0 & \cellcolor{gray!10} 0 & \cellcolor{gray!10} 0 & \cellcolor{gray!10} 0 & \cellcolor{gray!10} 0 &  
            \cellcolor{gray!10} 0 & \cellcolor{gray!10} 0 & \cellcolor{gray!10} 0 & \cellcolor{gray!10} 2 & \cellcolor{gray!10} 0 & \cellcolor{gray!10} 0 \\
           CSI-Cobot & $20$ ($23$-$11$) & 3 - 3 & 0 & 0 & 3 & 0 & 0 & 0 & 0 & 0 & 0 &  0 &  0 &  0 &  0 &  0\\
            \cellcolor{gray!10} DAISY &  
            \cellcolor{gray!10} $26$ (45-31) & \cellcolor{gray!10} 15 - 5 & \cellcolor{gray!10} 0 & \cellcolor{gray!10} 0 & \cellcolor{gray!10} 9 & \cellcolor{gray!10} 0 & \cellcolor{gray!10} 0 & \cellcolor{gray!10} 3 & \cellcolor{gray!10} 0 & \cellcolor{gray!10} 0 & 
            \cellcolor{gray!10} 3 & \cellcolor{gray!10} 5 & \cellcolor{gray!10} 0 & \cellcolor{gray!10} 4 & \cellcolor{gray!10} 0 & \cellcolor{gray!10} 0\\
            DPA & $26$ (28-25) & 3 - 2 & 0 & 0 &  2 & 0 & 0 & 1 & 0 & 0 & 0 &  0 &  0 &  0 &  0 &  0\\
            \cellcolor{gray!10} DressAssist & \cellcolor{gray!10} 31 (54-42) &  \cellcolor{gray!10} 0 - 0 & \cellcolor{gray!10} 0 & \cellcolor{gray!10} 0 & \cellcolor{gray!10} 0 & \cellcolor{gray!10} 0 & \cellcolor{gray!10} 0 & \cellcolor{gray!10} 0 & \cellcolor{gray!10} 0 & 
            \cellcolor{gray!10} 0 & \cellcolor{gray!10} 0 & \cellcolor{gray!10} 0 & \cellcolor{gray!10} 0 & \cellcolor{gray!10} 0 & \cellcolor{gray!10} 0 & \cellcolor{gray!10} 0\\
            SafeSCAD &  $28$ ($29$-$20$) & 1 - 8  & 0 & 0 & 0 & 0 & 1 & 0 & 0 & 0 & 0 & 0 & 0 & 0 & 0 & 0\\
            \bottomrule
        \end{tabular}
    }
    \vspace{-0.1in}
\end{table*}

\subsection{RQ1}

\boldparagraph{Models and methods}
N-TSs were asked to review the capabilities relation inferred by \SAN on nine real-world case-studies. 
The N-TS group included an ethicist, a lawyer, a philosopher, and a psychologist. 
The cases were taken from the repository of normative requirements~\cite{feng-et-al-24}: 
(1) ALMI~\cite{Hamilton-et-al-22}: a system assisting elderly or disabled users in a monitoring/advisory role and with everyday tasks; 
(2) ASPEN~\cite{aspen-23}: an autonomous agent dedicated to forest protection, providing both diagnosis and treatment of various tree diseases; 
(3) AutoCar~\cite{autocar-20}: a system that implements emergency-vehicle priority awareness for autonomous vehicles; 
(4) BSN~\cite{Gil-et-al-21}: a healthcare system detecting emergencies by continuously monitoring the patient’s health status; 
(5) DressAssist~\cite{Jevtic-et-al-19,townsend-et-al-2022}: an assistive and supportive system used to dress and provide basic care for those in need; 
(6) CSI-Cobot~\cite{Stefanakos-et-al-22}: a system ensuring the safe integration of industrial collaborative robot manipulators; 
(7) DAISY~\cite{daisy-22}: a sociotechnical AI-supported system that directs patients through an A\&E triage pathway (our running example); 
8) DPA~\cite{Amaral-et-al-22}: a system to check  compliance of data processing agreements against the General Data Protection Regulation; 
(9) SafeSCAD~\cite{calinescu2021maintaining}: a driver attentiveness management system to support safe shared control of autonomous vehicles.
We have selected these cases because they have been used in  existing work~\cite{feng-et-al-24} which examined them for well-formedness issues.  However, the authors of \cite{feng-et-al-24}assumed independence of all capabilities referenced in the rules.  Our goal is to determine whether capturing the relations between capabilities is important in practice.

\boldparagraph{Results}
We consider semantic relations to be \emph{correct} (TP) if they were classified as correct by the majority of the N-TSs.  The others are considered \emph{spurious} (FP).  
For each case, we report the type of semantic relation that has been captured as well as the new relevant well-formedness issues identified. 
The results are shown in Tbl.~\ref{tab:SAN-eval}.

\SAN inferred the total of 103 semantic relations for all but one (DressAssist) cases, and the N-TSs classified 53 of them as correct.  The classification took between 4 minutes (BSN) and  30 minutes (DPA).  The number of correct relations added per case ranged between 19 (ASPEN) and one (BSN and SafeSCAD). 
The majority of the added relations (27) involved contradicting events, that is, there were 27 pairs of events whose expected effects were mutually exclusive.
The second most common relation type added (13) was the one involving mutual exclusiveness between two measures.
We observed that a higher number of events and measures did not lead to a promortionally higher number of relations between them.
We identified 13 new well-formedness issues, all of which have been clasisfied as relevant by the N-TSs. 
Out of them, six were instances of vacuous conflicting rules, meaning that triggering them would inevitably lead  to a conflict!  The remaining seven instances represented redundant requirements.
To summarize, capturing semantic relations using \SAN enabled us to identify new relevant WFIs, answering {\bf RQ1}.

\subsection{RQ2}

\boldparagraph{Models and methods}
We conducted a controlled experiment involving two groups of non-technical stakeholders (N-TSs) across two real-world cases: (1) Tabiat~\cite{tabiat}, a smartphone application (and sensors) that records symptoms and keeps doctors updated on patients' chronic obstructive pulmonary disease conditions; (2) Casper~\cite{Moro-et-al-18}, a socially assistive robot designed to help people living with dementia by supporting activities of daily living at home and improving their quality of life.
The first group of stakeholders, consisting of a philosopher and a psychologist, was tasked with eliciting SLEEC requirements for the Tabiat case using \approach. 
The second group, consisting of a lawyer and a medical doctor, did the same, but without any tool support.
For each case-study, we provided stakeholders with a detailed description of the system's capabilities and its operating environment and conducted the following experiments:

(i) Group 1 elicited requirements without guidance for the Tabiat case and validated them manually. We refer to this experiment as \emph{adhoc elicitation}. This experiment was chosen to provide a baseline.

(ii) Group 2 elicited requirements for the Casper case following Stage I guidance of \approach and validated them manually. We refer to this experiment as \emph{systematic-elicitation}. This was done to evaluate the impact of having a structured approach to requirements elicitation. 

(iii) Group 1 and Group 2 elicited requirements for the Casper and Tabiat cases, respectively, following the overall \approach approach (which includes guided elicitation and automated validation). We refer to these experiments as \emph{\approach-elicitation-validation}. This experiment was aimed to measure how the systematic elicitation and the automation affect 
the quality of the requirements.

The  elicited requirements obtained in each experiment  are compared based on the size of the requirements set (including the number of events and measures) and number of well-formedness issues. 
To facilitate the comparison, we used \SAN to extract relations from the manually elicited requirements (\emph{adhoc-elicitation} and \emph{systematic-elicitation}), and then asked the stakeholders to filter them.

    \begin{table*}[t] 
        \caption{\small \approach compared to common practice. We record: true positive (TP) and false negative (FP) in the extracted capabilities relation (rel.); the number of new issues identified (after capturing the semantic relations) for vacuous conflict (v-conf.), situational conflict (s-conf.), redundancy (redund.), restrictiveness (restrict.), and insufficiency (insuff.); and the number of iterations for each WFI type.}
        \label{tab:RAINCOAT-eval}
        \centering
        \scalebox{0.8}{
            \begin{tabular}{r l c ||c | c c c c c | c c c c c}
                \toprule
                \multirow{2}{*}{experiments} & \multirow{2}{*}{cases} & 
                rules &  relations & \multicolumn{5}{c}{\WFI} & \multicolumn{5}{|c}{Number of iterations}  \\
                \cline{5-9} \cline{10-14}
                & & (evnt.$-$ msr.) & (TP - FP) & \multirow{1}{*}{v-conf.} & \multirow{1}{*}{s-conf.} & \multirow{1}{*}{redund.} & \multirow{1}{*}{insuffi.} & \multirow{1}{*}{restrict.} & \multirow{1}{*}{v-conf.} & \multirow{1}{*}{s-conf.} & \multirow{1}{*}{insuffi.} & \multirow{1}{*}{restrict.} & \multirow{1}{*}{redund.}\\ \toprule
                \emph{adhoc-elicitation} &Tabiat & $19$ ($27-13$) & 13 - 20 & 0 & 0 & 0 & \multicolumn{2}{c|}{not applicable}  & \multicolumn{5}{c}{not applicable} \\ \midrule
                 \emph{systematic-elicitation}&
                 Casper & $57$ ($59-24$) &  11 - 18 & 22 & 2 & 12 & \multicolumn{2}{c|}{not applicable}  & \multicolumn{5}{c}{not applicable} \\
                 \midrule
               \approach-&  Tabiat & $28$ ($37-18$) &  17 - 21 & 0 &  0 &  0 &  0 &  0 &   1 &   3 & 3 &  0 & 0\\
                \emph{elicitation-validation} & Casper  & $26$ ($38-14$) & 23 - 22 & 0 & 0 & 0 & 0 & 0 & 0 &   0 &  1 & 0 &  1 \\
                \bottomrule
            \end{tabular}
        }
        \vspace{-0.15in}
    \end{table*}

\boldparagraph{Results}
Tbl.~\ref{tab:RAINCOAT-eval} reports  the number of requirements, events, and measures, the correctly identified semantic relations (TP), the incorrectly identified ones (FP), and the number of different \WFI types
for the three experiments.

The experimental results show that the requirements elicited through the systematic approach are more comprehensive. i.e., result in a set covering more system capabilities and SLEEC principles compared to \emph{adhoc-elicitation}.  In the latter case,  
the stakeholders were guided by the SLEEC principles and applied them to a selection of  high-level system capabilities, but they did not explore all potential mappings with the complete set of system capabilities.  
We observed that \emph{adhoc-elicitation} was free from conflicts and redundancies, whereas
\emph{systematic-elicitation} had 34 WFIs, including 24 conflicts and 12 redundancies. The high number of resulting requirements in the \emph{systematic-elicitation}, 57, rendered
manual validation infeasible.
In contrast, \emph{adhoc-elicitation} focused on a smaller set of requirements for fewer system capabilities, but each with distinct and independent responses.

Comparing \emph{systematic-elicitation} with \emph{approach-elicitation-validation}, we observe
that the former yielded a significantly larger number of requirements.
Since \emph{systematic-elicitation} had to rely on manual analysis, the experimental results
revealed 
many redundant and conflicting requirements.
Moreover, we observed that when evaluating the insufficiency and restrictiveness of requirements elicited in \emph{approach-elicitation-validation}, the stakeholders gained confidence that they had elicited `enough' requirements to prevent SLEEC harms, ensuring that the system remained usable without overly restricting its main functionalities. The same stakeholders expressed regret over not implementing this  step during  \emph{systematic-elicitation}. For the \emph{approach-elicitation-validation}, on average, the stakeholders converged within three iterations and resolved all instances of WFI, which required  at least double the amount of time spent in \emph{systematic-elicitation}. 

We conclude that in our experiments \emph{systematic-elicitation} is crucial to elicit a comprehensive set of requirements that encompass all system capabilities. Although integrating it with automated validation (\emph{approach-elicitation-validation}) incurred a time cost, we conclude that it was essential for preventing the elicitation of an inconsistent set of requirements. Moreover, this combined approach, implemented in \approach, effectively empowered non-technical stakeholders to produce a well-defined set of requirements without compromising the elicitation process,
answering 
{\bf RQ2}.

\vspace{-0.05in}
\subsection{Threats to validity}
For the experiments answering both {\bf RQ1} and {\bf RQ2}, 
(1) the non-technical stakeholders are co-authors of this paper. We mitigated this threat by separating the authors into those who participated in developing the approach and those who evaluated it, and ensured that a complete separation between them was maintained throughout the entire lifecycle of the project.
(2)  Using ChatGPT for extracting semantic relations has expected risks, such as potential misinterpretation and lack of traceability. To mitigate these, we involved stakeholders in the process, asking them to review and either accept or reject the proposed relations. Although this step is time-consuming, it is still faster than a capturing the relations manually.

For the experiments answering {\bf RQ2}, 
(3) the two control groups in our study included stakeholders from diverse professional backgrounds: one group consisted of a philosopher and a psychologist, and the other of a lawyer and a medical doctor. To address potential bias due to their differing expertise, we avoided making comparisons based on profession-specific requirements, focusing only on the well-formedness of the overall set of elicited requirements.
(4) Our second set of experiments were limited to only two cases, which could potentially narrow the scope of our conclusions. However, we mitigated this by selecting real-world cases from different systems (i.e., a smartphone application and a robot), developed by stakeholders with diverse areas of expertise. 

%% file: Sections/related-work.tex
\section{Related Work}
\label{sec:relatedwork}



Requirement engineering, spanning from elicitation to validation, poses  significant challenges due to requirement uncertainties and potential misinterpretations by stakeholders. To address these challenges, ontology-driven techniques have been explored, aiming to capture semantic relations among the concepts articulated in the requirements documentation~\cite{ BencharquiOntologybasedRS, Diamantopoulos2018EnhancingRR}. These approaches emphasize hierarchical relations (e.g., ``is class of'', ``is instance of'') between concepts~\cite{Farfeleder2005MKSO, Murugesh2015J}, facilitating requirements specification through relation analysis. While knowledge graphs require substantial efforts to build the entire language elements~\cite{knowledge-graphs, Kejriwal2019DomainSpecificKG}, our approach  prioritizes semantic relations, considering the domain knowledge as an input to enhance the efficacy of the elicitation process, without claiming linguistic completeness.

Recently, LLM-based techniques have emerged to streamline requirements engineering tasks, particularly in bridging the gap between natural language requirements and their specifications \cite{arora2023advancing}. Investigations into the potential role of ChatGBT in the elicitation process have explored both its benefits and limitations \cite{ronanki2023investigating}. In this paper, LLM assistance aims to extract relations among language elements of \sleeckeyword{events} and \sleeckeyword{measures}, which we term the capability relations of the system for a sound reasoning and analysis, rather than constructing a requirements specification from natural language documents~\cite{Fantechi2023GPS}.

%% file: Sections/conclusion.tex
\section{Conclusion}
\label{sec:conclusion}
We have introduced a novel technique which leverages LLMs to capture common sense and bridge the gap between manually and automatically analyzing requirements. This is achieved by extracting semantic relations between the abstract representations of system capabilities in the normative requirements. 
These relations are then used to enrich the automated-reasoning techniques for eliciting and analyzing the consistency and coherence of normative requirements. 
The importance of capturing such relationships has been demonstrated by the identification of 13 new well-formedness issues (\WFI) across nine existing case studies. 
The effectiveness and usability of our  approach have been demonstrated on two real world studies with a total 64 relationships captured and 130 requirements which were elicited by stakeholders with diverse backgrounds.
%
Our technique for extracting semantic relationships could be improved to reduce the number of false positives. In the future, we plan to explore enhancements, e.g., by refining the prompting strategies or fine-tuning the LLM to better perform the task. Our proposed approach lacks automatic support for the debugging and resolution of \WFI concerns.  Semi-automated generation of patches for \WFI{s} is left for future work.

\section*{Acknoledgements}
The authors would like to thank Daniyal Liaqat for providing the Tabiat system specification for our experiments.
This project has received funding from the RAI UK international partnership project `Disruption Mitigation for Responsible AI', the UKRI project EP/V026747/1 `Trustworthy Autonomous Systems Node in Resilience', and the Royal Academy of Engineering Grant No CiET1718/45.

%% file: Sections/Appendix/appendix.tex
\subsection{Semantics of \dsl}\label{app:semantics}
\begin{figure*}
\begin{tabular}{lcl}
            \centering
            $\fos \models 
            \hyp{\event_a}{\event_b}$ & iff & $\forall (\eventocc{i}, \measureassign{i}, \timestamp{i}) \in \fos \cdot \event_a \in \eventocc{i} \Rightarrow \event_b \in \eventocc{i}$ \\
            $\fos \models 
            \con{\event_a}{\event_b}$ & iff & $\forall (\eventocc{i}, \measureassign{i}, \timestamp{i}) \in \fos \cdot \neg (\event_a \in \eventocc{i} \wedge \event_b \in \eventocc{i})$ \\
            $\fos \models 
            \hb{\event_a}{\event_b}$ & iff & $\forall (\eventocc{i}, \measureassign{i}, \timestamp{i}) \in \fos \cdot  (\event_b \in \eventocc{i} \Rightarrow \exists (\eventocc{j}, \measureassign{j}, \timestamp{j}) \in \fos \cdot (\event_a \in \eventocc{j} \wedge i > j))$ \\
              $\fos \models 
            \eq{\event_a}{\event_b}$ & iff & $\fos \models \hyp{\event_a} \wedge {\event_b} \models \hyp{\event_b}{\event_a} $ \\
              $\fos \models 
 \imply{\prop_a}{\prop_b}$ & iff & $\forall (\eventocc{i}, \measureassign{i}, \timestamp{i}) \in \fos \cdot \measureassign{i}(\prop_a \Rightarrow \prop_b) = \top$ \\
             $\fos \models  \me{\prop_a}{\prop_b}$ & iff & $\forall (\eventocc{i}, \measureassign{i}, \timestamp{i}) \in \fos \cdot \measureassign{i}(\prop_a \wedge \prop_b) = \bot$ \\
             $\fos \models  \ops{\prop_a}{\prop_b}$ & iff & $\forall (\eventocc{i}, \measureassign{i}, \timestamp{i}) \in \fos \cdot \measureassign{i}(\prop_a \iff \prop_b) = \bot$ \\
            $\fos \models  \eq{\prop_a}{\prop_b}$ & iff & $\forall (\eventocc{i}, \measureassign{i}, \timestamp{i}) \in \fos \cdot \measureassign{i}(\prop_a \iff \prop_b) = \top$ \\
            $\fos \models  \fob{\prop_a}{\prop_b}$ & iff & $\forall (\eventocc{i}, \measureassign{i}, \timestamp{i}) \in \fos \cdot \event_a \in \eventocc{i} \Rightarrow \measureassign{i}
            (\prop_{j}) = \bot$ \\
             $\fos \models \ind{\prop_a}{\prop_b}$ & iff & $\forall (\eventocc{i}, \measureassign{i}, \timestamp{i}) \in \fos \cdot \event_a \in \eventocc{i} \Rightarrow \measureassign{i}
            (\prop_{j}) = \top$ \\
            $ \fos \models \wuntil{\event_a}{\prop_b}{\event_c}$ & iff & \multicolumn{1}{c}{\makecell{$\forall (\eventocc{i}, \measureassign{i}, \timestamp{i}) \cdot \event_a \in \eventocc{i} \Rightarrow (\exists (\eventocc{j}, \measureassign{j}, \timestamp{j}) \cdot \event_c \in \eventocc{j} \wedge \forall k \in [i,j) \cdot 
            \measureassign{j}(\prop_b) = \top)
            $ \\ $ \vee \forall (\eventocc{j}, \measureassign{j}, \timestamp{j}) \in \fos \cdot j\ge i \Rightarrow  \measureassign{j}(\prop_b) = \top)$ }}\\
            $ \fos \models \wfor{\event_a}{\prop_b}{\term}$ & iff & \multicolumn{1}{c}{\makecell{$\forall (\eventocc{i}, \measureassign{i}, \timestamp{i}) \cdot \event_a \in \eventocc{i} \Rightarrow 
             \forall (\eventocc{j}, \measureassign{j}, \timestamp{j}) \in \fos \cdot (\timestamp{j} \in [\timestamp{i}, \timestamp{i} + \measureassign{i}(\term))) \Rightarrow \measureassign{i}(\prop_b) = \top)$ }}\\

\end{tabular}
\caption{The semantics of DSL relations defined on a trace $\fos = (\eventocc{1}, \measureassign{1}, \timestamp{1}), (\eventocc{2}, \measureassign{2}, \timestamp{2}),$ $\ldots  (\eventocc{n}, \measureassign{n}, \timestamp{n})$.}
\label{tab:rsemantic}
\end{figure*}

The semantics of \dsl is described over  \textit{traces}, which are \textit{finite} sequences of \textit{states} $\fos = (\eventocc{1}, \measureassign{1}, \timestamp{1}), (\eventocc{2}, \measureassign{2}, \timestamp{2}),$ $\ldots  (\eventocc{n}, \measureassign{n}, \timestamp{n})$. For every time point $i \in [1, n]$, (1) $\eventocc{i}$ is a set of events that occur at $i$; (2) $\measureassign{i}: \measures \rightarrow \mathbb{N}$ assigns every measure in $\measures$ to a concrete value at time point $i$, and (3) $\timestamp{i}: \mathbb{N}$ captures the value of time point $i$ (e.g., the second time point can have the value 30, for 30 sec). We assume that the time values in the trace are strictly increasing (i.e., $\timestamp{i} < \timestamp{i+1}$ for every $i \in [1, n-1]$). Given a measure assignment $\measureassign{i}$ and a term $\term$, let $\measureassign{i}(\term)$ denote the result of substituting every measure symbol $\measure$ with $\measureassign{i}(\measure)$. Since $\measureassign{i}(\term)$ does not contain free variables, the substitution results in a natural number. 
 Similarly, given a proposition $\prop$, we say that $\measureassign{i} \models \prop$ if $\prop$ is evaluated to $\top$ after substituting every term $\term$ with $\measureassign{i}(\term)$.  
Let a trace $\fos = (\eventocc{1}, \measureassign{1}, \timestamp{1}) \ldots (\eventocc{n}, \measureassign{n}, \timestamp{n})$ be given.

\italicparagraph{Positive obligation} A positive obligation $\within{\event}{\term}$ is \textit{fulfilled subject to time point $i$}, denoted as $\fos \models_{i} \within{\event}{\term}$, if there is a time point $j \ge i$ such that $e \in \eventocc{j}$ and $\timestamp{j} \in [\timestamp{i}, \timestamp{i} + \measureassign{i}(\term)]$. That obligation is \emph{violated at time point $j$}, denoted as $\fos \not\models_{i}^{j} \within{\event}{\term}$, if $\timestamp{j} = \timestamp{i} + \measureassign{i}(\term)$, and for every $j'$ such that $j \ge j' \ge i$, $\event$ does not occur ($\event \not\in \eventocc{j'}$).  

\italicparagraph{Negative obligation} A negative obligation $\within{\notS \; \event}{\term}$ is fulfilled subject to time point $i$, denoted as  $\fos \models_{i} \within{\notS \; \event}{\term}$, if for every time point $j$ such that $\timestamp{j} \in [\timestamp{i}, \timestamp{i} + \measureassign{i}(\term)]$, $e \not\in \eventocc{j}$. 
The negative obligation is violated at time point $j$, denoted as $\fos \not\models_{i}^{j} \within{\notS \; \event}{\term}$, if (1) $\timestamp{j} \in [\timestamp{i}, \;\timestamp{i} + \measureassign{i}(\term)]$, (2) $\event$ occurs ($\event \in \eventocc{j}$), and (3) for every $j \ge j' \ge i$, $\event$ does not occur ($\event \not\in \eventocc{j'}$). 

\italicparagraph{Conditional obligation} A conditional obligation $\prop \Rightarrow \obg$ is fulfilled subject to time point $i$, denoted as $\fos \models_{i} (\prop \Rightarrow \obg)$, if $\prop$ does not hold at time point $i$ (that is, $\measureassign{i}(\prop) = \bot$) or the obligation is fulfilled ($\fos \models_{i} \obg$).  Moreover, $\prop \Rightarrow \obg$  is violated at time point $j$, denoted as $\fos \not\models_{i}^{j} (\prop \Rightarrow \obg)$, if  $\prop$ holds at $i$ ($\measureassign{i}(\prop) = \top$) and $\obg$ is violated at $j$ ($\fos \not\models_{i}^{j} \obg$).

\vspace{-0.02in}
\italicparagraph{Obligation chain} A chain $\cobg^+_1 \otherwise \cobg^+_2 \ldots \cobg_m$ is fulfilled subject to time point $i$, denoted as $\fos \models_{i} \cobg^+_1 \otherwise$ $\cobg^+_2 \ldots \cobg_m$, if (1) the first obligation is fulfilled ($\fos \models_{i} \cobg^+_1$) or (2) 
there exists a time point $j \ge i$ such that $\cobg^+_1$ is \textit{violated} (i.e., $\fos \not\models_{i}^{j} \cobg_1$) and the rest of the obligation chain (if not empty) is fulfilled at time point $j$ ($\fos \models_{j} \cobg^+_2 \otherwise \ldots \cobg_m$).

\vspace{-0.02in}
\italicparagraph{Rule} A rule $\rulesyntax{\event \; \andS \; \prop}{\dobg}$ is fulfilled in $\fos$, denoted as $\fos \models \rulesyntax{\event \; \andS \; \prop}{\dobg}$, if for every time point $i$, 
where event $\event$ occurs ($\event \in \eventocc{i}$) and $\prop$ holds ($\measureassign{i}(\prop) = \top$), the obligation chain is fulfilled subject to time point $i$ ($\fos \models_{i} \dobg$). A \emph{trace fulfills a rule set} $\ruleset$, denoted as $\fos \models \ruleset$, if it fulfills every rule in the set (i.e., for every $\srule \in \ruleset$,  $\fos \models \srule$). 


\begin{definition}[Behaviour defined by $\ruleset$]\label{def:behav}
    Let a rule set  $\ruleset$ be given. The accepted \emph{behaviour} defined by $\ruleset$, denoted as $\lang{\ruleset}$, is the largest set of traces such that every trace $\fos$ in $\lang{\ruleset}$ respects  $\ruleset$, i.e., $\fos \in \lang{\ruleset}$, $\fos \models \ruleset$.
\end{definition}
    On top of the usual rules, \dsl also uses \emph{\facts} as an utility construct  describing the test scenarios for rules. A \emph{fact} (see Tab.~\ref{tab:syntax}) $\factsyntax{\event \wedge \prop}{(\dobg \mid \notS \dobg)}$ asserts the existence of an event $\event$ under condition $\prop$ while an obligation chain $\dobg$ is either satisfied  or violated. \facts~ can be used to describe undesirable behaviors, denoted as \textit{concerns}, which should be blocked by \dsl rules. Alternatively, \facts\; can also be used to describe functional goals, denoted as \textit{purposes}, which should be allowed by \dsl rules.

\subsection{Semantics of SLEEC relations}~\label{app:relsem}
Let $\fos$ be a sequence of states $(\eventocc{1}, \measureassign{1}, \timestamp{1}), (\eventocc{2}, \measureassign{2}, \timestamp{2}),$ $\ldots  (\eventocc{n}, \measureassign{n}, \timestamp{n})$ where $\eventocc{i}$, $\measureassign{i}$, and $\timestamp{i}$ are the occurrence of events, valuation of measures, and the clock time of state $i$, respectively. The semantics of SLEEC relationships is defined on $\fos$ in Fig.~\ref{tab:rsemantic}.